\documentclass[universe,article,accept,pdftex,moreauthors]{Definitions/mdpi}

\usepackage{amsmath}    
\usepackage{amsfonts}   
\usepackage{mathtools}  
\usepackage{tensor}     
\usepackage{graphicx}   
\usepackage{caption}     
\usepackage{subcaption}  
\usepackage{accents}     

\renewcommand\arraystretch{1.1} 
\newcommand{\be}{\begin{equation}}
\newcommand{\ee}{\end{equation}}
\def\bea{\begin{eqnarray}}
\def\eea{\end{eqnarray}}

\newcommand{\cH}{\ensuremath{\mathcal{H}}}
\firstpage{1} 
\pubvolume{1}
\issuenum{1}
\articlenumber{0}
\pubyear{2025}
\copyrightyear{2025}
\externaleditor{Firstname Lastname} 
\datereceived{30 September 2025} 
\daterevised{25 October 2025} 
\dateaccepted{4 November 2025} 
\datepublished{ } 
\hreflink{https://doi.org/} 

\Title{Observational Tests of the Conformal Osculating Barthel--Kropina Cosmological Model}

\TitleCitation{Observational Tests of the Conformal Osculating Barthel--Kropina Cosmological Model}

\Author{Himanshu Chaudhary 
 $^{1}$, Rattanasak Hama $^{2}$, Tiberiu Harko $^{1,3,}$*, Sorin V. Sabau $^{4,5}$ and Shibesh Kumar Jas Pacif $^{6,7}$}

\AuthorNames{Himanshu Chaudhary, Rattanasak Hama, Tiberiu Harko, Sorin V. Sabau, and Shibesh Kumar Jas Pacif}

\AuthorCitation{Chaudhary, H.; Harko, T.; Hama, R.; Sabau, S. V.; Pacif, S. K. J.}

\address{%
$^{1}$ \quad Department of Physics, Babeș-Bolyai University, Kogălniceanu Street 1, Cluj Napoca 400084, Romania; himanshu.chaudhary@ubbcluj.ro\\

$^{2}$ \quad Faculty of Science and Industrial Technology, Prince of Songkla University, Surat Thani Campus,\linebreak   Surat Thani 84000, Thailand; rattanasak.h@psu.ac.th \\

$^{3}$ \quad Astronomical Observatory, 19 Ciresilor Street, Cluj-Napoca 400487, Romania \\

$^{4}$ \quad School of Biological Sciences, Department of Biology, Tokai University, Sapporo 005-8600, Japan; sorin@tokai.ac.jp \\

$^{5}$ \quad Graduate School of Science and Technology, Physical and Mathematical Sciences, Tokai University,\linebreak   Sapporo 005-8600, Japan, \\

$^{6}$ \quad Pacif Institute of Cosmology and Selfology (PICS), Sagara, Sambalpur 768224, Odisha, 
 India;
director@pacif-ics.com \\

$^{7}$ \quad Research Center of Astrophysics and Cosmology, Khazar University, 41 Mehseti Street,\linebreak   Baku AZ1096, 
 Azerbaijan 
}

\corres{Correspondence:tiberiu.harko@aira.astro.ro}

\abstract{We consider detailed cosmological tests of dark energy models obtained from the general conformal transformation of the Kropina metric, representing an $(\alpha,\beta)$-type Finslerian geometry. In particular, we restrict our analysis to the osculating Barthel--Kropina geometry. The Kropina metric function is defined as the ratio of the square of a Riemannian metric $\alpha$ and of the one-form $\beta$. In this framework, we also consider the role of the conformal transformations of the metric, which allows us to introduce a family of conformal Barthel–Kropina theories in an osculating geometry. The models obtained in this way are described by second-order field equations, in the presence of an effective scalar field induced by the conformal factor. The generalized Friedmann  equations of the model are obtained by adopting for the Riemannian metric $\alpha$ the Friedmann--Lemaitre--Robertson--Walker representation. In order to close the cosmological field equations, we assume a specific relationship between the component of the one-form $\beta$  and the conformal factor. With this assumption, the cosmological evolution is determined by the initial conditions of the scalar field {and a single free parameter $\gamma $ of the model}. The conformal Barthel--Kropina cosmological models are compared against several observational datasets, including Cosmic Chronometers, Type Ia Supernovae, and Baryon Acoustic Oscillations, using a Markov Chain Monte Carlo (MCMC) analysis, {which allows the determination of $\gamma$}. A comparison with the predictions of standard $\Lambda$CDM model is also performed. {Our results indicate that the conformal osculating Barthel--Kropina model can be considered as a successful, and simple, alternative to standard cosmological models.} }


\begin{document}


\section{Introduction}

The investigation of the role of the conformal transformations in physics may provide an important avenue for the understanding of gravitational phenomena, of the elementary particle physics, and of their relationship. The idea of the conformal transformations (rescalings) was first proposed by Weyl \cite{Weyl1,Weyl2,Weyl3}, in his attempt for finding a unified theory of gravitation and electromagnetism.  The conformal transformations were called Weyl gauge transformations, and they are presently a standard method in elementary particle physics.  Gauge field theories are the basic theoretical tools of our present-day approach to field theory, and they allow deep insights into the properties of the elementary particles. In gravitational theories, the Weyl gauge transformations are called conformal transformations.  The invariance of physical laws under conformal transformations is called conformal invariance. Many important geometrical quantities are also conformally invariant. Many important equations of physics, including the Maxwell equations, satisfy the requirement of local scale invariance. 

The important role of conformal transformations was emphasized by T. Hooft in~\cite{Hooft1,Hooft2}, who pointed out that conformal symmetry may be an exact symmetry of nature,  which is broken spontaneously.  Conformal symmetry could be as important as the Lorentz symmetry of the laws of nature,  and it may open some new directions for the understanding of the physics of the Planck scale. In a theory of the gravitational interaction proposed in \cite{Hooft2}, it is assumed that local conformal symmetry is an exact, but spontaneously broken symmetry of nature. The conformal part of the metric is interpreted as a dilaton field. The theory has intriguing implications, and it suggests that black holes are topologically trivial, regular solitons, without singularities,  firewalls, or horizons.

The role of the conformal transformations was also explored by Penrose \cite{Pen1,Pen2,Pen3,Pen4}, in the framework of a cosmological model called Conformal Cyclic Cosmology (CCC). The starting point of this model is the observation that when the de Sitter accelerating stage, induced by the presence of the positive cosmological constant $\Lambda$, ends, the spacetime is conformally flat, and space-like. This geometric structure  coincides with the geometry of the initial boundary of the Universe after the Big Bang. Moreover, in the Conformal Cyclic Cosmology model, it is suggested  that the Universe consists of eons, time-oriented manifolds, possessing space-like null infinities.  Conformally invariant gravitational theories were investigated in \cite{Man1,Man2,Man3, Gh1,Gh2, Gh3,Gh4,Ha1,Ha2,Ha3,Gh5,Ha4,Ha5,Gh6,Gh7,Ha6,Gh8}.

In 1918, in the year Weyl presented his extension of Riemann geometry, another important geometric theory was also introduced. This is the Finsler geometry \cite{Fin}, representing another  important generalization of Riemann geometry. Even from a purely mathematical perspective, Finsler geometry is ``... just Riemannian geometry without the quadratic restriction'' \cite{Chern}. We will still consider Finsler geometry as a generalization of Riemann geometry. The Finsler geometry was already anticipated by Riemann \cite{Riem}, who, in a general space, introduced  a geometric structure given by  $ds = F \left(x^1,...,x^n;dx^1,...,dx^n = F(x,dx)\right)$. According to this definition,  for a nonzero $y, y\neq 0$, the function $F(x,y)$, the Finsler metric function, must be a positive function defined on the tangent bundle TM.  $F(x,y)$ must also satisfy the important requirement of homogeneity of degree one in $y$, which implies the condition $F(x,\lambda dx) = \lambda F(x,dx)$, where $\lambda$ is a positive constant. The case  $F^2 = g_{ij}(x)dx^idx^j$ leads to the limiting case of the Riemann geometry \cite{Bao, Fin1}.

An important class of Finsler spaces is represented by the Kropina spaces \cite{Kr1,Kr2}. The Kropina spaces are Finsler spaces of the type $(\alpha,\beta)$, in which the Finsler metric function $F$ is defined as $F = \alpha ^2/\beta$, where $\alpha$ denotes a Riemannian metric, $\alpha (x,y) = \left(g_{IJ}y^Iy^J\right)^{1/2}$, and $\beta (x,y) = A_I(x)dy^I$ is a one-form. The properties of the Kropina spaces have been considered, from a mathematical perspective, in \cite{Fin2,Fin3,Fin4,Fin5,Fin6,Fin7,Fin8}. A significant simplification of the mathematical approach  can be obtained  through the use of the theory of the osculating Riemann spaces of Finsler geometries, introduced in \cite{O1,O2}. In this approach, one replaces a complex Finsler geometric object with a simpler mathematical object, represented, for example, by a Riemann metric. Thus, using the osculating formalism, a simpler mathematical and geometrical description can be obtained. In the specific case of the Kropina metric, one can take the field $Y (x)$ as $Y(x) = A(x)$, and then the  A-osculating Riemannian manifold is defined as $\left(M,\hat{g}(x,A(x))\right)$. Moreover, one can associate to this mathematical structure the Barthel connection, representing  the Levi--Civita connection of the Riemann metric $\hat{g}_{IJ}(x) = \hat{g}_{IJ}(x,A(x))$.

One of the most important scientific developments of the 20th century is represented by Einstein’s theory of general relativity (GR). Its remarkable success is mainly due to the introduction of the  geometric description of the gravitational interaction \cite{book1,book2}. GR is a fundamental theory of matter, gravity, and space--time, matter, and of their interaction, giving an extremely precise description of the gravitational effects at the level of the Solar System. Unfortunately, when extended to very small, and very large astrophysical or cosmological scales,  GR is confronted with several challenges, coming mostly from quantum field theory and cosmology. 

First of all, a large number of recent cosmological observations have raised important concerns about the possibility of considering GR as the theoretical foundation of cosmology. The recent acceleration of the Universe \cite{acc1,acc2,acc3} can be explained remarkably well by reintroducing in the Einstein field equations the cosmological constant $\Lambda$ \cite{L}, together with another mysterious, and yet undetected,  matter component, called dark matter, a pressureless and cold constituent of the Universe. In recent years, the $\Lambda$CDM ($\Lambda$ Cold Dark Matter) cosmological model has become a standard approach for the analysis and interpretation of the observational cosmological data. In this direction, the $\lambda$CDM model proved to be very successful. However, there are a number of important questions suggesting that $\Lambda$CDM may be considered as only representing a first-order approximation of a more general model, not yet known \cite{Rev}. 

One of the basic challenges the $\Lambda$CDM model faces is the lack of a consistent physical and theoretical background, related to the absence of a convincing explanation or description of the cosmological constant. Presently, no convincing physical or geometrical interpretation of $\Lambda$ is known. Another weakness of the $\Lambda$CDM model is related to the nature of dark matter. After many years of intensive observational and experimental efforts, no positive detection of the particles associated  to dark matter has been yet recorded in both terrestrial experiments, or astrophysical observations. 

\textls[-15]{The technological advances in the field of cosmology, which led to the significant increase in the precision of observations, revealed another important weakness of the $\Lambda$CDM standard model. Significant deviations have been found between the Hubble  expansion rates obtained from  the low redshift (local) measurements  and those measured by the Planck satellite experiment using the Cosmic Microwave Background Radiation (CMBR). The differences obtained from the different determinations of the values of the present-day Hubble constant $H_0$ are usually called the Hubble tension, and, if confirmed, it could represent a paradigmatic crisis in present-day cosmology 
 \cite{T1,T2,T3,T4,T5,T6}. The difference in the numerical values of $H_0$ obtained by the Planck satellite, $H_0 = 66.93 \pm 0.62$ km/s/Mpc 
 \cite{T5,T6}, and the values of $H_0 = 73.24 \pm 1.74$ km/s/Mpc \cite{T3} inferred by the SH0ES collaboration, is greater than 3$\sigma$~\cite{T6}. If it indeed exists, the Hubble tension is a strong indicator of the need of developing new gravitational theories, and for the necessity of replacing the $\Lambda$CDM model with an alternative and more realistic one. }

General relativity also faces important challenges on a theoretical ground. The Big Bang singularity, and generally the presence of singularities in the theory, which are extremely important for the understanding of the origin of the Universe, and of its very early evolution, is still unexplained by general relativity, and the $\Lambda$CDM cosmology. Moreover,  GR cannot describe the extremely high density phases of matter, in the presence of extremely strong gravitational fields, as is the case for black holes. From another theoretical perspective,  very little progress has been made, if any, in the understanding of the quantum properties of gravity, including the quantization of geometry, spacetime, and gravity~\cite{E1}. Therefore, in the absence of a quantum description of gravity, as yet, GR cannot be considered as a fundamental physical theory, similar to the theories describing elementary particle interactions, and their properties. 

One possibility to obtain a solution of these fundamental problems is to consider generalized theories of gravity, which contain GR as a particular case, corresponding to a weak field limit. There are many attempts for constructing alternative theoretical approaches to GR. These novel theories are obtained by introducing and developing different physical and mathematical approaches (for detailed reviews of modified gravity theories, and of their cosmological and astrophysical implications, see \cite{E2,E3,E4,E5}.

One of the interesting extensions of standard general relativity are represented by gravitational theories based on the extensive use of Finsler geometry. Finsler-type gravitational theories, as well as the cosmological models obtained using these geometrical structures, represent important alternatives to standard cosmology. Generally, they can provide a geometric explanation of dark energy, and even of dark matter. Many studies have been devoted to the applications of the Finsler geometry in cosmology and gravitational physics. These studies have led  to a new understanding and to a new geometric perspective on the cosmological evolution, dark energy, and dark matter \cite{F1,F2,F2a, F3, F4, F5, F6,F7,F8, F9, F10, F11, F12,F13, F14,F15}. For a recent review of the cosmological applications of the Finsler geometry, see \cite{F16}. 

In particular, a special type of Finsler geometry, the Barthel--Kropina spaces, have been also investigated extensively from the point of view of their cosmological applications \cite{B1,B2,B3}. The Barthel–Kropina cosmological approach is generally based on the introduction of a Barthel connection in an osculating $(\alpha, \beta)$-type Finsler geometry.  The Barthel connection has the important property that it is the Levi--Civita connection of a Riemannian metric. By assuming that the background Riemannian metric is of the Friedmann–Lemaitre–Robertson–Walker type, one can obtain the generalized Friedmann equations of the Barthel–Kropina models. These equations show that an effective geometric dark energy component can be generated within the framework of the Barthel--Kropina geometries, having an effective, geometric-type energy density and pressure, respectively. To fully solve the cosmological models, generally, one must impose an equation of state for the dark energy.

A specific Barthel--Kropina type cosmological model was considered in \cite{B3}. The model was obtained from the general conformal transformation of the $(\alpha, \beta )$ Kropina metric, and the possibilities of obtaining conformal theories of gravity in the osculating Barthel–Kropina geometric framework were investigated. A family of conformal Barthel–Kropina gravitational field theories in an osculating geometry with second-order field equations were introduced. The models depend on the properties of the conformal factor, whose presence leads to the appearance of an effective scalar field of geometric origin in the gravitational field equations. The cosmological implications of the theory were investigated in detail by assuming a specific relation between the component of the one-form of the Kropina metric and the conformal factor. The cosmological evolution is thus determined by the initial conditions of the scalar field and a free parameter of the model. 

It is the goal of the present paper to extend the investigations initiated in \cite{B3}, by considering a detailed analysis of the cosmological tests, including a full comparison with observational data of this dark energy model that are considered in detail. To constrain the conformal Barthel–Kropina model parameters, and the values of the scalar field, we use 57 Hubble data points, the Pantheon Supernovae Type Ia data sample, and the BAO (Baryonic Acoustic Oscillations) measurements. The statistical analysis is performed using Markov Chain Monte Carlo (MCMC) simulations. A detailed comparison with the standard $\Lambda$CDM model is also performed, with the Akaike information criterion (AIC), and the Bayesian information criterion (BIC) is used as the two model selection tools. The statefinder diagnostics consisting of jerk and snap parameters, and the $Om(z)$ diagnostic tools, are also considered for the comparative study of the conformal Barthel–Kropina and $\Lambda$CDM cosmologies. Our results indicate that for certain values of the model parameters, the Barthel–Kropina dark energy model produces a good description of the observational data, and thus it can be considered a viable alternative of the $\Lambda$CDM model, by also alleviating some of the theoretical problems standard cosmology is facing.

The present paper is organized as follows. The theoretical foundations of the conformal Barthel--Kropina-type gravitational theories are introduced in Section~\ref{sect1}, where the generalized Friedmann equations are also written down. The cosmological tests of the theory are considered, for a few values of the model parameter, in Section~\ref{sect2}. A summary and a discussion of the cosmological implications of the theory is presented in Section~\ref{sect3}. We conclude the results of our investigations in Section~\ref{sect4}. 

\section{Conformal Barthel–Kropina Cosmology}\label{sect1}

In the present section, by following the approach initiated in \cite{B3},  we will introduce the fundamentals of the conformal transformations in Riemann and Barthel--Kropina geometries, respectively, and we will present the generalized Friedmann equations of the osculating conformal Barthel--Kropina geometry.

\subsection{Conformal Transformations in Riemann Geometry} 

The conformal transformation of a Riemannian metric $g_{ij}(x)$, defined as 
\cite{book2}:
\be
\tilde{g}_{ij}(x)=\Omega ^2(x)g_{ij}(x)=e^{2\sigma(x)}g_{ij}(x),
\ee
where $\sigma (x)$ is an arbitrary function of the coordinates $x$, and plays an important role in both mathematics and physics. 
In Riemannian geometry, the conformal transformations are defined on the space--time manifold $M$, and  $\tilde{g}_{ij}$ denotes the conformally transformed Riemannian metric. 
The Christoffel symbols of $\tilde{g}_{ij}$ and $g_{ij}$, and  $\tilde{\gamma}^i_{jk}$ and $\gamma^i_{jk}$, respectively, are then related as 
\begin{equation}\label{Lema1}
\begin{split}
\tilde{\gamma}^i_{jk}&=\gamma^i_{jk}+\delta^i_j\sigma_k+\delta^i_k\sigma_j-\sigma^i g_{jk},\\
\end{split}
\end{equation}
where 
\be
\sigma_j:=\frac{\partial\sigma(x)}{\partial x^j}, \sigma^i=g^{ij}\sigma_j.
\ee

For the transformation law  of the covariant derivative for any $X=X^i\frac{\partial}{\partial x^i}$, we obtain the expression \cite{B3}:
\be
\tilde {\nabla }_{X}Y=\nabla _{X}Y+d\sigma (X)Y+d\sigma (Y)X-g(X,Y)\nabla \sigma,
\ee
where we have denoted
\be
\nabla\sigma=\sigma^i\frac{\partial}{\partial x^i}, d\sigma(x)=\frac{\partial\sigma(x)}{\partial x^i}X^i.
\ee

\subsubsection*{Matter and Energy--Momentum Tensor}

\textls[-15]{In standard general relativity, the matter action can be generally written in the form~\cite{book1,book2}}:
\be
S_m=\int{L_m(g,\psi)\sqrt{-g}d^4x},
\ee
where  $L_m$ is the matter Lagrangian, which we consider to be a function of the metric tensor $g$, and of the matter fields  $\psi$, which can be of bosonic or fermionic nature. We assume that under conformal transformations, the matter Lagrangian transforms according to the rule
\be\label{CL}
\tilde{L}_m=e ^{-4\sigma (x)}L_m,
\ee
and thus the conformally transformed action takes the form \cite{B3}:
\bea
\tilde{S}_m&=&\int{\tilde{L}_m\sqrt{-\tilde{g}}d^4x}=\int{e^{-4\sigma (x)}L_m e^{4\sigma (x)}\sqrt{-g}d^4x}=S_m.
\eea

Thus, the action of the ordinary baryonic matter is invariant under the conformal transformations (\ref{CL}). This important result indicates that since the baryonic matter action is an invariant quantity, in all conformally related frames, it can be described by the same~expression.

The matter energy--momentum tensor $T_{I J}$ is defined as \cite{book2}:
\be
T_{I J}=\frac{2}{\sqrt{-g}}\frac{\delta }{\delta g^{I J}}\left(\sqrt{-g}L_m\right),
\ee
and, after a conformal transformation of the metric it becomes \cite{B3}:
\be
\tilde{T}_{I J}=e ^{-2 \sigma (x)} T_{I J}.
\ee
The 
 trace $\tilde{T}=\tilde{T}_I^I$ of the baryonic matter energy--momentum tensor transforms in a conformal transformation as $\tilde{T}=e ^{-4\sigma (x)}T$, where $T=T_I^I$.

\subsection{The osculating Barthel--Kropina cosmological model}

The Kropina metric function, a specific Finsler type $(\alpha,\beta)$ metric, is defined according to \cite{B3}:
\be
F=\frac{\alpha^2}{\beta}=\frac{g_{IJ}(x)y^Iy^J}{A_I(x)y^I},\ I,J=\{0,1,2,3\}.
\ee
We also denote  $y_I:=g_{IJ}y^J$. The fundamental tensor associated to the metric is given by~\cite{B3}:
\bea\label{eq_Kropina Hessian}
\hat{g}_{IJ}(x,y)&=&\frac{2\alpha^2}{\beta^2}g_{IJ}(x)+\frac{3\alpha^4}{\beta^4}A_IA_J-\frac{4\alpha^2}{\beta^3}(y_IA_J+y_JA_I)+\frac{4}{\beta^2}y_Iy_J,
\eea

For the tensor components $g_{IJ}$ of the Riemannian metric $\alpha $, we consider the expression
\be
(g_{IJ}(x))=\begin{pmatrix}
1 & 0 & 0 & 0\\
0 & -a^2(x^0) & 0 & 0\\
0 & 0 & -a^2(x^0) & 0\\
0 & 0 & 0 & -a^2(x^0)
\end{pmatrix},
\ee
which describes the homogeneous and isotropic, flat Friedmann--Lemaitre. For the one-form $\beta$, we adopt the expression  $\beta=A_I(x)y^I=A_0(x)y^0$,
where
\be
(A_I(x))=(A_0,0,0,0)=(a(x^0)\eta(x^0),0,0,0),
\ee
is a covariant vector field defined on the base manifold  $M$.

We also consider the preferred direction
\be
Y=Y^I\frac{\partial}{\partial x^I}=A^I\frac{\partial}{\partial x^I},
\ee
where $ A^I:=g^{IJ}A_J$. For the case of the FLRW metric, we have \cite{B3}:
\be
(Y^I)\equiv(A^I)=(Y_I)=(A_I)=(a(x^0)\eta(x^0),0,0,0),
\ee
where $\eta \left(x^0\right)$ is an arbitrary function of the time coordinate $x^0$. 

We can now obtain
\begin{equation}
\beta\vert_{y=A}=[a(x^0)\eta(x^0)]^2,
\ee
and
\be
\begin{split}
(h_{IJ})\vert_{y=A}&=\begin{pmatrix}
0 & 0 & 0 & 0\\
0 & -a^2(x^0) & 0 & 0\\
0 & 0 & -a^2(x^0) & 0\\
0 & 0 & 0 & -a^2(x^0)
\end{pmatrix},
\end{split}
\end{equation}
where $h_{IJ}:=g_{IJ}(x)-\frac{y_I}{\alpha}\frac{y_J}{\alpha}$, $y_I:=g_{IJ}(x)y^J$.

By substituting the above relations  in \eqref{eq_Kropina Hessian}, we obtain the osculating Riemannian metric of the Barthel--Kropina cosmological model as \cite{B3}:
\begin{equation}\label{eq_osculating_Kropina Hessian}
\begin{split}
&\hat{g}_{IJ}(x)=\hat{g}_{IJ}(x,y=A)
=\begin{pmatrix}
\frac{1}{a^2\eta^2} & 0 & 0 & 0\\
0 & -\frac{2}{\eta^2} & 0 & 0\\
0 & 0 & -\frac{2}{\eta^2} & 0\\
0 & 0 & 0 & -\frac{2}{\eta^2}
\end{pmatrix}
\end{split}
\end{equation}

The Einstein field equations are given by 
\be
\hat{G}_{00}=\frac{8\pi G}{c^4}\hat{g}_{00}\rho c^2, 
\ee
and 
\be
\hat{G}_{ii}=-\frac{8\pi G}{c^4}\hat{g}_{ii}p, 
\ee
where $\rho$ and $p$ denote the matter energy density, and pressure, respectively. Hence, the system of the generalized Friedmann equations in the osculating Barthel--Kropina geometry take the form \cite{B3}:
\be\label{Fr1}
\frac{3(\eta')^2}{\eta^2}=\frac{8 \pi G}{c^4}\frac{1}{a^2\eta ^2}\rho c^2,
\ee
\be\label{Fr2}
a^2\left[-3(\eta')^2+2\eta\eta''+2\cH\eta\eta'\right]=\frac{8\pi G}{c^4}p,
\ee
where $\cH=a'\left(x^0\right)/a\left(x^0\right)$ is the generalized Hubble function. {In the following, by a prime, we denote the derivative with respect to the coordinate $x^0=ct$, so that $a'=da\left(x^0\right)/dx^0$.} By substituting the term $-3\left(\eta '\right)^2$ and with the use of Equation~(\ref{Fr1}), it follows that  Equation~(\ref{Fr2}) can be simplified to
\be\label{Fr3}
2a\eta \frac{d}{dx^0}\left(\eta 'a\right)=\frac{8\pi G}{c^4}\left(\rho c^2+p\right).
\ee

The cosmological implications of this model have been investigated in detail in \cite{B1} and \cite{B2}, respectively. 

\subsection{The Conformal Osculating Barthel--Kropina Model---the Generalized Friedmann Equations}

For a general $(\alpha,\beta)$ metric with metric function $F=F(\alpha,\beta)$, we introduce the conformal transformation \cite{B3}:
\be\label{Cftrans}
\tilde{F}(x,y):=e^{\sigma(x)}F(x,y)=\tilde{F}(\tilde{\alpha},\tilde{\beta}),
\ee
which is an $(\tilde{\alpha},\tilde{\beta})$ metric, where
\be\label{Cftrans1}
\tilde{\alpha}=e^{\sigma(x)}\alpha,\ \tilde{\beta}=e^{\sigma(x)}\beta.
\ee

The fundamental tensor of $\tilde{F}$ is given by the Hessian
\be
\tilde{g}_{IJ}:=\frac{1}{2}\frac{\partial^2\tilde{F}^2}{\partial y^I\partial y^J}.
\ee

For the case of the Kropina metric, its conformal transform is given by \cite{B3}:
\be
\tilde{F}:=e^{\sigma(x)}\frac{\alpha^2}{\beta}=\frac{\tilde{\alpha}^2}{\tilde{\beta}},
\ee
where 
\be
\tilde{\alpha}=e^{\sigma(x)}\alpha, \tilde{\beta}=e^{\sigma(x)}\beta.
\ee

The conformally transformed osculating Riemannian metric is obtained in the form
\be\label{14}
\hat{\tilde{g}}_{IJ}(x)=e^{2\sigma(x)}\hat{g}_{IJ}(x),
\ee
where $\hat{g}_{IJ}(x)$ is given by Equation~\eqref{eq_osculating_Kropina Hessian}. Explicitly, the metric $\hat{g}_{IJ}(x)$ has the expression
\begin{equation}\label{conf_metr}
\begin{split}
&\hat{\tilde{g}}_{IJ}(x)=e^{2\sigma (x)}
\begin{pmatrix}
\frac{1}{a^2\eta^2} & 0 & 0 & 0\\
0 &- \frac{2}{\eta^2} & 0 & 0\\
0 & 0 & -\frac{2}{\eta^2} & 0\\
0 & 0 & 0 & -\frac{2}{\eta^2}
\end{pmatrix}
.
\end{split}
\end{equation}

\subsection{The Generalized Cosmological Evolution Equations}

We assume now that the Einstein gravitational field equations are given in the conformal osculating Barthel--Kropina as \cite{B3}:
\be\label{Ein}
\hat{\tilde{G}}_{IJ}=\frac{8\pi G}{c^4}\hat{\tilde{T}}_{IJ},
\ee
where by $\hat{\tilde{T}}_{IJ}$, we have denoted the matter energy--momentum tensor in the conformal frame. We further assume that the thermodynamic properties of the baryonic matter in the conformal Barthel--Kropina cosmological models can be described by the conformal energy density $\hat{\tilde{\rho}} c^2$, and the conformal thermodynamic pressure $\hat{\tilde{p}}$ only. We also introduce another important assumption; namely, we postulate the existence of a coordinate frame comoving with matter. Therefore, the energy--momentum tensor of the baryonic matter takes, in the conformal frame, the form
 \be
\hat{\tilde{T}}_{I}^{J}=\begin{pmatrix}
\hat{\tilde{\rho}} c^2 & 0 & 0 & 0\\
0 & -\hat{\tilde{p}} & 0 & 0\\
0 & 0 & -\hat{\tilde{p}} & 0\\
0 & 0 & 0 & -\hat{\tilde{p}}
\end{pmatrix},
\ee
and
 \be
\hat{\tilde{T}}_{IJ}=e^{-2\sigma (x)}\begin{pmatrix}
\frac{e^{2\sigma (x)}}{a^2\eta ^2}\hat{\tilde{\rho}} c^2 & 0 & 0 & 0\\
0 & \frac{2e^{2\sigma (x)}}{\eta ^2}\hat{\tilde{p}} & 0 & 0\\
0 & 0 & \frac{2e^{2\sigma (x)}}{\eta ^2}\hat{\tilde{p}} & 0\\
0 & 0 & 0 & \frac{2e^{\sigma (x)}}{\eta ^2}\hat{\tilde{p}}
\end{pmatrix},
\ee
respectively.

The homogeneity and isotropy of the space--time implies that all physical and geometrical quantities can depend only on the time coordinate $x^0$. Moreover,  we assume that the conformal factor has the form \cite{B3}:
\be
\sigma (x)=\phi \left(x^0\right)+\gamma _1x+\gamma _2y+\gamma _3z,
\ee
where $\gamma_i$, $i=1,2,3$ are arbitrary constants. For this choice, the Einstein field equations produce $\hat{\tilde{G}}_{ij}
=-\gamma_i\gamma_j=0$, which leads to $\gamma _i=0$, $i=1,2,3$. Therefore, without any loss in generality, we chose the conformal factor as $\sigma (x)=\phi \left(x^0\right)$, and thus, in the following, we consider only time-dependent conformal transformations of the Kropina metric \cite{B3}.

\subsection{The generalized Friedmann equations}

By taking into account all the above results, it follows that the generalized Friedmann equations, in the conformal osculating Barthel--Kropina cosmology, take the form \cite{B3}:
\be\label{B1}
\frac{3(\eta')^2}{\eta^2}=\frac{8\pi G}{c^2}\frac{1}{a^2\eta ^2}\hat{\tilde{\rho}}+3\left(\phi '\right)^2-6\frac{\eta '}{\eta }\phi ',
\ee
and
\bea\label{B2}
\hspace{-1.0cm}&&\frac{2}{\eta^2}[-3(\eta')^2+2\eta\eta'\cH+2\eta\eta'']
=\frac{16\pi G}{c^4}\frac{1}{a^2\eta ^2}\hat{\tilde{p}}
-4\left[\phi ''+\frac{1}{2}\left(\phi '\right)^2\right]
+\left(\frac{\eta'}{\eta}-\cH\right)\phi',
\eea
respectively. After eliminating the term $-3\left(\eta '\right)^2/\eta ^2$ between Equations~(\ref{B1}) and (\ref{B2}), we obtain the equation
\bea
2\frac{1}{a\eta}\frac{d}{dx^0}\left(a\eta '\right)&=&\frac{8\pi G}{c^4}\frac{1}{a^2\eta ^2}\left(\hat{\tilde{\rho}}c^2+\hat{\tilde{p}}\right)-\left(\phi ''-\left(\phi '\right)^2\right)
-\frac{11}{4}\frac{\eta '}{\eta} \phi'
-\frac{1}{4}\phi ' \cH.
\eea

We consider now the possibility of the description of the dark energy as a geometric effect in the Barthel--Kropina cosmological model \cite{B3}. In the limit $\eta\rightarrow 1/a$, and $\phi =0$, we recover the standard general relativistic model without a cosmological constant. The deviations  from general relativity, and standard Riemannian geometry can be considered as a small variation of $\eta$, depending on the conformal factor. Thus, we introduce a cosmological model in which $\eta$ is given by \cite{B3}:
\be\label{eta}
\eta =\frac{e^{\gamma \phi}}{a},
\ee
where $\gamma $ is a constant. 

The generalized Friedmann equations (\ref{B1}) and (\ref{B2}) of  the conformal osculating Barthel--Kropina model take the form
\bea\label{58}
\hspace{-0.5cm}3\cH ^2&=&\frac{8\pi G}{c^2}e^{-2\gamma \phi}\hat{\tilde{\rho}}+3\left(1-2\gamma-\gamma ^2\right)\left(\phi '\right)^2+6(1+\gamma)\phi ' \cH 
=\frac{8\pi G}{c^2}e^{-2\gamma \phi}\hat{\tilde{\rho}}+\hat{\tilde{\rho}}_\phi,
\eea
and
\bea\label{59}
2\cH '+3\cH ^2&=&-\frac{8\pi G}{c^4}e^{-2\gamma \phi}\hat{\tilde{p}}+2(1+\gamma)\phi '' 
-\left(1-\frac{1}{2}\gamma -\gamma ^2\right)\left(\phi '\right)^2
+(1+4\gamma)\phi '\cH\nonumber\\
&=&-\frac{8\pi G}{c^4}e^{-2\gamma \phi}\hat{\tilde{p}}-\hat{\tilde{p}}_\phi,
\eea
respectively, where we have introduced the notations
\be
\hat{\tilde{\rho}}_\phi=3\left(1-2\gamma-\gamma ^2\right)\left(\phi '\right)^2+6(1+\gamma)\phi ' \cH,
\ee
and
\be
\hat{\tilde{p}}_\phi=-2(1+\gamma)\phi '' +\left(1-\frac{1}{2}\gamma -\gamma ^2\right)\left(\phi '\right)^2-(1+4\gamma)\phi '\cH,
\ee
respectively.

The total conservation equation for matter and the conformally induced scalar field can be obtained from the generalized Friedmann equations in the form
\bea\label{63}
&&\frac{8\pi G}{c^4}\left[\hat{\tilde{\rho}}'+3\cH\left(\hat{\tilde{\rho}}+\frac{\hat{\tilde{p}}}{c^2}\right)-2\gamma \phi ' \hat{\tilde{\rho}} \right]e^{-2\gamma \phi}
+\hat{\tilde{\rho}}'_\phi+3\cH \left(\hat{\tilde{\rho}}_\phi+\hat{\tilde{p}}_\phi\right)=0.
\eea

 Equation~(\ref{63}) can be split into two independent balance equations, one for matter, and the second for the conformal scalar field, respectively, which take the form
\be\label{64}
\hat{\tilde{\rho}}'+3\cH\left(\hat{\tilde{\rho}}+\frac{\hat{\tilde{p}}}{c^2}\right)-2\gamma \phi ' \hat{\tilde{\rho}} =0,
\ee
and
\be\label{65}
\hat{\tilde{\rho}}'_\phi+3\cH \left(\hat{\tilde{\rho}}_\phi+\hat{\tilde{p}}_\phi\right)=0,
\ee
respectively. {We would like to point out that the splitting represented by Equations~(\ref{64}) and (\ref{65}) is a supplementary assumptions introduced in order to simplify the mathematical structure of the model, and it represents a particular choice for the evolution of the scalar field. Models in which the two terms in the general conservation Equation (\ref{63}) are not zero, or  one is the negative with respect to the other, are also possible. From a purely mathematical point of view, the considered splitting can be thought of as a constraint imposed on the evolution of thermodynamic parameters of the the matter and scalar field, respectively. Moreover, this assumption allows us to obtain a consistent and systematic solution of the generalized Friedmann equations, without the need of considering further physical or cosmological inputs.}

From Equation~(\ref{65}), we obtain the time evolution of the conformal scalar field as
\bea\label{66}
\hspace{-0.5cm}&&6\left(1-2\gamma -\gamma ^2\right)\phi ''+3\left(4-\frac{13}{2}\gamma -4\gamma ^2\right)\cH\phi' 
 +6(1+\gamma)\cH '+3(2\gamma +5)\cH ^2=0.
\eea

Thus,  the basic equations describing the cosmological dynamics in the conformal Barthel--Kropina model are given by Equations~(\ref{58}), (\ref{59}), and (\ref{66}), respectively \cite{B3}.

 Equations~(\ref{58}), (\ref{59}), and (\ref{66}) produce, for $\cH '$ and $\phi ''$, the expressions
\bea
\cH '&=&\frac{1}{4}\left( \gamma ^{2}+2\gamma -1\right) \frac{8\pi G}{c^{2}}\hat{\tilde{\rho}}
e^{-2\gamma \phi }
-\frac{1}{8}\left( 8\gamma ^{4}+29\gamma ^{3}+10\gamma
^{2}-29\gamma +8\right) \left( \phi ^{\prime }\right) ^{2}\nonumber\\
&&-\frac{1}{4}\left(
2\gamma ^{2}+7\gamma +5\right) \cH^{2}
+\frac{3}{8}\left( 4\gamma ^{3}+13\gamma
^{2}+7\gamma -6\right) \cH\phi ^{\prime },\
\eea
and
\bea
\phi ^{\prime \prime }&=&\frac{1}{4}(\gamma +1)\frac{8\pi G}{c^{2}}\hat{\tilde{\rho}}
e^{-2\gamma \phi }
-\frac{1}{8}\left( 8\gamma ^{3}+21\gamma ^{2}+5\gamma
-8\right) \left( \phi ^{\prime }\right) ^{2}\nonumber\\
&&-\frac{1}{4}(2\gamma +5)\cH^{2}+%
\frac{1}{8}\left( 12\gamma ^{2}+27\gamma +2\right) \cH\phi ^{\prime },
\eea
respectively.

We introduce now, in the generalized Friedmann equations instead of the coordinate $x^0=ct$, the time $t$, and instead the Hubble function $\cH$, the time-dependent Hubble function $H=\dot{a}/a$, where by a dot we have denoted the derivative with respect to the cosmological time $t$. Hence, we have $\cH=H/c$. 

To simplify the mathematical formalism, we define a set of dimensionless variables $\left(h,\tau,r_m\right)$, which are given by 
\be
H=H_0h, \tau =H_0t, \hat{\tilde{\rho}}=\frac{3H_0^2}{8\pi G}r_m,
\ee
where $H_0$ is the present-day value of the Hubble function. Hence, the full set of the evolution equations of the conformal osculating Barthel--Kropina cosmological model is obtained in the form
\be\label{70}
\frac{dr_m}{d\tau}+3hr_m=2\gamma \frac{d\phi}{d\tau}r_m,
\ee
\begin{eqnarray}\label{71}
\frac{dh}{d\tau } &=&\frac{3}{4}\left( \gamma ^{2}+2\gamma -1\right)
r_{m}e^{-2\gamma \phi }  
-\frac{1}{8}\left( 8\gamma ^{4}+29\gamma +10\gamma ^{2}-29\gamma
+8\right) \left( \frac{d\phi }{d\tau }\right) ^{2}  \nonumber \\
&&-\frac{1}{4}\left( 2\gamma ^{2}+7\gamma +5\right) h^{2}
+\frac{3}{8}\left(
4\gamma ^{3}+13\gamma ^{2}+7\gamma -6\right) h\frac{d\phi }{d\tau },
\end{eqnarray}%
and
\begin{eqnarray}\label{72}
\hspace{-0.5cm}\frac{d^{2}\phi }{d\tau ^{2}} &=&\frac{3}{4}(\gamma +1)r_{m}e^{-2\gamma \phi
}  
-\frac{1}{8}\left( 8\gamma ^{3}+21\gamma ^{2}+5\gamma -8\right) \left(
\frac{d\phi }{d\tau }\right) ^{2}  \nonumber \\
\hspace{-0.5cm}&&-\frac{1}{4}(2\gamma +5)h^{2}+\frac{1}{8}\left( 12\gamma ^{2}+27\gamma
+2\right) h\frac{d\phi }{d\tau },
\end{eqnarray}%
respectively. {In the following, we consider that the Universe is filled only with pressureless dust, and thus we will take the matter pressure as zero. Moreover, no other matter, radiation, or dark matter contributions are added, and dark energy and dark matter are considered together as a geometric effect generated by the presence of a Finslerian-type geometry. Hence, the only equation of state that is needed is the equation of state of the pressureless matter, $p_m=0$.
}

After integrating Equation~(\ref{70}) for the matter density under the assumption of pressureless dust, we obtain
\be
r_m(\tau)=r_{m0}^{\prime}\frac{e^{2\gamma \phi}}{a^3}=r_{m0}^{\prime}\frac{\eta ^2}{a},
\ee
where $r_{m0}^{\prime}$ is an arbitrary constant of integration. 

Next, in order to obtain a form of the evolution equations that allow an easy comparison with the observational data, we introduce the redshift variable $z$ defined as $1+z=1/a$, {with $a_0=a(0)=1$ representing the present-day value of the scale factor, according to the standard conventions of present-day cosmology}. 

\textls[-25]{{In order to numerically solve the system of three cosmological evolution Equations (\ref{70})--(\ref{72}), we will proceed in two steps. First, we reformulate the second-order system as a first-order dynamical system, by introducing the new variable $u=d\phi/d\tau$, giving $d^2\phi/d\tau ^2=du/d\tau$, and $(d\phi/d\tau)^2=u^2$. Secondly, we will reformulate the evolution equations in the redshift space, by replacing the cosmological time $\tau$ as independent variable with the redshift $z$. Consequently, the time derivative can be replaced by the derivative with respect to $z$ according to the relation}
\be
\frac{d}{d\tau}=-(1+z)h(z)\frac{d}{dz}.
\ee   
}

Therefore,  the cosmological evolution equations in the redshift space take the form~\cite{B3}:
\be\label{F1}
-(1+z)h\frac{d\phi}{dz}=u,
\ee
\be\label{F2}
(1+z)\frac{dr_m}{dz}-3r_m=2\gamma (1+z)\frac{d\phi}{dz}r_m,
\ee
\begin{eqnarray}\label{F3}
-(1+z)h\frac{dh}{dz} &=&\frac{3}{4}\left( \gamma ^{2}+2\gamma -1\right)
r_{m}e^{-2\gamma \phi } 
-\frac{1}{8}\left( 8\gamma ^{4}+29\gamma ^{3}+10\gamma ^{2}-29\gamma
+8\right) u^{2}  \nonumber \\
&&-\frac{1}{4}\left( 2\gamma ^{2}+7\gamma +5\right) h^{2}
+\frac{3}{8}\left(
4\gamma ^{3}+13\gamma ^{2}+7\gamma -6\right) hu,
\end{eqnarray}
and
\begin{eqnarray}\label{F4}
-(1+z)h\frac{du}{dz} &=&\frac{3}{4}(\gamma +1)r_{m}e^{-2\gamma \phi }
-\frac{1}{8}\left( 8\gamma ^{3}+21\gamma ^{2}+5\gamma -8\right) u ^{2}  \nonumber \\
&&-\frac{1}{4}(2\gamma +5)h^{2}+\frac{1}{8}\left( 12\gamma ^{2}+27\gamma
+2\right) hu, \nonumber\\
\end{eqnarray}
respectively. The system of Equations (\ref{F1})--(\ref{F4}) must be integrated with the initial conditions $h(0)=1$, $\phi (0)=\phi_0$, $u(0)=u_0$, and $r_m(0)=r_{m0}$, respectively. At the present time, the initial value of the matter density $r_m(0)$ is related to the integration constant $r_{m0}^{\prime}$ by the relation $r_{m0}=r_{m0}^{\prime}e^{\gamma \phi_0}$. 

\section{Observational Tests of the Conformal Barthel--Kropina\linebreak   Cosmological Models}\label{sect2}

{In this section, we present the constraints on the parameters of the conformal osculating Barthel–Kropina (COBK) cosmological model, whose theoretical formulation has been introduced in the previous sections. To do this, we use the cosmological model described by the set (\ref{F1})–(\ref{F4}) of differential equations. We then perform the MCMC analysis to constrain the parameters of the COBK Model. The system of equations is solved numerically using the initial conditions $\phi(0)=\phi_0$, $u(0)=u_0$, $r_m(0)=r_{m0}$, and $h(0)=1$, respectively, while $\gamma$ is assumed to be a free parameter.}

\subsection{Methodology and Datasets}

To constrain the parameters of the {COBK model}, we first numerically solve the system of differential equations of the conformal osculating Barthel–Kropina cosmological model given by Equations (\ref{F1})–(\ref{F4}), by specifying the appropriate initial conditions, and by integrating over the redshift $z$. The system of equations is solved for the redshift interval $z\in[0,3]$ using the \textit{solve\_ivp} routine with the Radau method, which is an implicit Runge–Kutta solver particularly well suited for stiff systems of differential equations. Then, we use the numerical solution obtained in this way, and apply the \textsc{Nested Sampling} {algorithm}, implemented with the \textsc{PyPolyChord} library, to constrain the parameters of the COBK Model
\endnote{\url{https://github.com/PolyChord/PolyChordLite}} \cite{Handley2015,Handley2015b}. 

Nested sampling is particularly well suited for cosmological inference, as it not only provides robust posterior distributions for parameter estimation, but also directly computes the Bayesian evidence, $\mathcal{Z}$. In this framework, the posterior probability distribution is expressed as $P(\theta|D, M) = \frac{\mathcal{L}(\theta)\,\pi(\theta)}{\mathcal{Z}}$,  where $\mathcal{L}(\theta)$ is the likelihood, $\pi(\theta)$ is the prior, and $\mathcal{Z}$ is the evidence obtained by marginalizing over the full parameter space, $\mathcal{Z} = \int \mathcal{L}(\theta)\,\pi(\theta)\, d\theta$.

\textls[-15]{Beyond parameter estimation, the evidence plays a key role in model selection. Competing cosmological models are compared using the Bayes factor, $B_{ab} \equiv \frac{\mathcal{Z}_a}{\mathcal{Z}_b}, \quad \ln B_{ab} \equiv \Delta \ln \mathcal{Z},$ which quantifies the statistical preference of one model over another. }

\textls[-25]{To interpret the strength of the evidence, we follow the Jeffreys’ scale \cite{Jeffreys}: $0 \leq |\Delta \ln \mathcal{Z}| < 1$ indicates inconclusive or weak support, $1 \leq |\Delta \ln \mathcal{Z}| < 3$ corresponds to moderate evidence, $3 \leq |\Delta \ln \mathcal{Z}| < 5$ suggests strong evidence, and $|\Delta \ln \mathcal{Z}| \geq 5$ is regarded as decisive in favor of the model with higher evidence.}

In our analysis, we employ \textsc{PyPolyChord} with 300 live points to ensure robust Bayesian evidence estimation. For visualization, we use the \texttt{getdist} package\endnote{\url{https://github.com/cmbant/getdist}} \cite{Lewis2025}, which provides marginalized posterior distributions and parameter correlation plots. This study incorporates multiple dataset combinations, including Baryon Acoustic Oscillation measurements, Type Ia supernovae, and Cosmic Chronometers, as detailed below.

\begin{itemize}
\item \textbf{Baryon Acoustic Oscillation: 
} 
We use the Baryon Acoustic Oscillation (BAO) measurements from over 14 million galaxies and quasars provided by the Dark Energy Spectroscopic Instrument (DESI) Data Release 2 (DR2)\endnote{\url{https://github.com/CobayaSampler/bao\_data}}\cite{Karim2025}. To constrain the cosmological parameters using BAO from DESI DR2, we compute three primary distance measures: the Hubble distance $D_H(z) = c/H(z)$, the comoving angular diameter distance $D_M(z) = c \int_0^z dz'/H(z')$, and the volume-averaged distance $D_V(z) = [z D_M^2(z) D_H(z)]^{1/3}$. These distances are expressed as ratios $D_H(z)/r_d$, $D_M(z)/r_d$, and $D_V(z)/r_d$ for direct comparison with the observed BAO data. Here, $r_d$ denotes the sound horizon at the drag epoch ($z_d \approx 1060$), defined as $r_d = \int_{z_d}^{\infty} c_s(z)/H(z)\, dz$, where $c_s(z)$ is the sound speed of the photon–baryon fluid. While the standard flat $\Lambda$CDM model predicts $r_d = 147.09 \pm 0.26$ Mpc \cite{Aghanim2020}, we treat $r_d$ as a free parameter in our analysis \cite{Pogosian2020,Jedamzik2021,Pogosian2024,Lin2021,Vagnozzi2023}.
\item \textbf{Type Ia supernova:}
We  use the Pantheon$^{+}$ (PP) CosmoSIS likelihood\endnote{\url{https://github.com/PantheonPlusSH0ES/DataRelease}} in our analysis, which accounts for both statistical and systematic uncertainties through a covariance matrix \cite{Conley2010}. This dataset includes 1,590 light curves from 1,550 Type Ia Supernovae (SNe Ia) spanning the redshift range $0.01 \leq z \leq 2.26$ \cite{Brout2022}. Light curves at $z < 0.01$ are excluded due to significant systematic uncertainties arising from peculiar velocities. In this analysis, we also marginalize over the parameter $\mathcal{M}$; for further details, see Equations (A9–A12) of \cite{Goliath2001}.

\item \textbf{Cosmic Chronometers:}
We also consider the Hubble measurements obtained through the differential age method. This technique relies on passively evolving massive galaxies, formed at redshifts $z \sim 2-3$, providing a direct, model-independent estimate of the Hubble parameter via the relation $\Delta z / \Delta t$ \cite{Jimenez}. In this analysis, we use the likelihood provided by Moresco on his GitLab repository\endnote{\url{https://gitlab.com/mmoresco/CCcovariance}}, which incorporates the full covariance matrix to account for both statistical and systematic uncertainties \cite{Moresco2018,Moresco2020}. This likelihood includes Hubble parameter measurements spanning the redshift range $0.179 \leq z \leq 1.965$ \cite{Moresco2012,Moresco2015,Moresco2016}.
\end{itemize}

To 
 constrain the parameters of the present cosmological model, we maximize the overall likelihood function, which is defined as $\mathcal{L}_{\text{tot}} = \mathcal{L}_{\text{BAO}} \times \mathcal{L}_{\text{SNe Ia}} \times \mathcal{L}_{\text{CC}}$.



\subsection{Comparing {the Conformal Osculating Barthel–Kropina Model} with the $\Lambda$CDM Model}\label{sect4}

In this subsection, we plot the Hubble function and its residuals with respect to redshift, after obtaining the corresponding mean values of the {COBK model. We then compare the results of the COBK model with the $\Lambda$CDM model and the CC measurements proposed in \cite{Moresco2012,Moresco2015,Moresco2016}.  This analysis allows us to show the compatibility of the COBK model with the $\Lambda$CDM model and the Hubble measurement datasets.}

\subsubsection*{Evolution of the Hubble Parameter and Hubble Residual}

To compare the {COBK model} against the $\Lambda$CDM model and the {Hubble measurements}, as a first step, we plot the Hubble function of the $\Lambda$CDM model using the following expression
$$
H_{\Lambda \text{CDM}}(z) = H_0 \sqrt{\Omega_{m0} (1 + z)^3 + \Omega_{\Lambda0}},
$$
where $H_0 = 67.8\, \mathrm{km\,s^{-1}\,Mpc^{-1}}$ and $\Omega_{m0} = 0.309$.

{We consider the COBK model as being defined by Equations (\ref{F1})–(\ref{F4}). The corresponding numerical solution is obtained using the initial conditions as obtained from the statistical analysis. The numerical solution for the Hubble function is denoted by $h(z)$, and it is then scaled by the factor $H_0$, yielding the final form of the Hubble function, $H(z) = H_0 h(z)$. Furthermore, the Hubble residual is defined as}
$$
{\Delta H(z) = H_{\text{Conformal Osculating Barthel–Kropina}}(z) - H_{\Lambda\text{CDM}}(z),}
$$
{where $H_{\text{Conformal Osculating Barthel–Kropina}}(z)$ denotes the Hubble parameter of the COBK model, and $H_{\Lambda\text{CDM}}(z)$ is the Hubble parameter predicted by the standard $\Lambda$CDM model. This comparison allows us to quantify the deviations of the COBK model from the standard $\Lambda$CDM predictions. By examining both the Hubble function $H(z)$ and the residual $\Delta H(z)$, we can assess how the COBK model fits the $\Lambda$CDM and Hubble measurements.}


\subsection{Cosmographic Analysis of {the Conformal Osculating Barthel–Kropina Model}, and of the  $\Lambda$CDM Model}\label{sect5}

Cosmography serves as a robust, model-independent tool to probe the dynamical features of the Universe by investigating the behavior of the cosmological observables with respect to the redshift. This approach does not rely on the underlying gravitational theory, or specific assumptions about dark energy components, but rather on the kinematic properties of the cosmic expansion. The formalism uses the Taylor expansion of the scale factor $a(t)$ around the present epoch, expressing the expansion history in terms of measurable quantities like the Hubble parameter $H(z)$, the deceleration parameter $q(z)$, and higher-order derivatives such as the jerk parameter $j(z)$, the snap parameter $s(z)$, and the lerk parameter $l(z)$ \cite{Cosmographic1,Cosmographic2,Cosmographic3,Cosmographic4}.

\subsubsection*{Deceleration Parameter $q(z)$ and Jerk Parameter $j(z)$}
The deceleration parameter is mathematically expressed as
$$
q(z) = \frac{1}{H^2(z)} \frac{dH(z)}{dz}(1+z) - 1,
$$
\textls[-15]{where $H(z)$ is the Hubble parameter at redshift $z$. A negative $q(z)$ implies accelerated expansion, whereas a positive value suggests a decelerating Universe. Two significant cosmographic markers derived from $q(z)$ include its present-day value $q_0 = q(z=0)$, representing the current expansion state, and the transition redshift $z_{\text{tr}}$, defined by $q(z_{\text{tr}}) = 0$, which marks the epoch where the Universe shifted from deceleration to acceleration.
The jerk parameter, capturing the rate of change of acceleration, is defined as \cite{jerksnap}}
$$
j(z)
= 1
- 2(1+z)\,\frac{1}{H(z)}\frac{dH(z)}{dz}
+ (1+z)^2\,\frac{1}{H(z)}\frac{d^2 H(z)}{dz^2}
+ (1+z)^2\left(\frac{1}{H(z)}\frac{dH(z)}{dz}\right)^2.
$$

In the standard $\Lambda$CDM cosmological model, $j(z)$ remains constant at $j(z) = 1$, independent of redshift. Deviations from this value indicate the presence of additional dynamical effects beyond the cosmological constant.


\subsection{Dimensionless Matter Density $r_m$, Conformal Factor $\phi$, Effective Energy Density, and Effective Pressure of the Scalar Field}\label{sect5}

The dimensionless matter density is a fundamental cosmological parameter that provides a key test of the consistency of different cosmological models. It accounts for the total matter content of the Universe, including both baryonic and dark matter, and its present-day value sets the initial conditions for the Universe’s cosmological evolution. While the cosmological models do not predict the current matter density directly, they describe how it evolves during the earlier stages of the Universe.

The conformal factor arises from conformal transformations in geometry and physics, which modify the metric in a way that preserves angles but not necessarily lengths. In practical terms, the conformal factor is a function that rescales the metric at every point, allowing shapes to stretch or shrink while keeping the local angle structure intact. This factor plays an important role in modeling the geometrical and physical properties of spacetime in various cosmological frameworks.

The effective dimensionless energy density of the conformal scalar field can be obtained from the first Friedmann equation as 
\begin{equation}
r_\phi (z)=3h^2(z)-e^{-2\gamma \phi (z)}r_m  (z).
\end{equation}

The effective pressure associated to the scalar field is given by
\begin{equation}
p_\phi (z)=2(1+z)h(z)\frac{dh}{dz}-3h^2(z). 
\end{equation}





\subsection{Model Selection, and Statistical Assessment of {the COBK Cosmological Model}}\label{sect6}

In this section, we use statistical metrics to evaluate the performance and complexity of the {COBK Model}. These tests quantify how well the models fit the observational data, compare them to $\Lambda$CDM, and determine whether the additional parameters significantly improve the fit without adding unnecessary complexity.

\subsubsection{Goodness of Fit}

First, we use the Chi-squared statistic, $\chi^2$, to evaluate the performance of the {COBK model}, as it quantifies the discrepancy between theoretical predictions and observational data. To account for models with different numbers of free parameters, we consider the reduced Chi-squared statistic given by
$$
\chi^2_{\text{red}} = \frac{\chi^2_{\text{tot}}}{\text{DOF}},
$$
where DOF is the number of data points minus the number of fitted parameters. Values of $\chi^2_{\text{red}} \approx 1$ indicate a good fit, significantly higher values suggest a poor fit, and much lower values may signal overfitting.

\subsubsection{Model Comparison Using Information Criteria}

Then, we use information criteria to evaluate both the goodness of fit and the complexity of the {COBK model} relative to $\Lambda$CDM. These criteria are based on the minimum Chi-squared value, $\chi^2_{\text{min}}$, and include the Akaike Information Criterion (AIC) and the Bayesian Information Criterion (BIC) \cite{Liddle,AIC1,AIC2,AIC3,BIC1}, defined as 
$$
\text{AIC} = \chi^2_{\text{min}} + 2 \mathcal{P}, \quad 
\text{BIC} = \chi^2_{\text{min}} + \mathcal{P} \ln(\mathcal{N}),
$$
where $\mathcal{P}$ is the number of free parameters, $\mathcal{N}$ is the total number of observational data points ($\mathcal{N} = 1618$ in our analysis), and $\chi^2_{\text{min}}$ represents the minimum Chi-squared achieved by the model. Both criteria penalize models with more parameters to avoid overfitting, with BIC generally applying a stronger penalty for larger datasets. For reference, the $\Lambda$CDM model has three free parameters, while the {COBK model has six free parameters.}

\subsubsection{Relative Comparison: $\Delta$AIC and $\Delta$BIC}

To compare the {COBK model} directly with $\Lambda$CDM, we compute the differences
$$
{\Delta \text{AIC} = \text{AIC}_{\text{Conformal Osculating Barthel–Kropina}} - \text{AIC}_{\Lambda \text{CDM}},}
$$
$$
{\Delta \text{BIC} = \text{BIC}_{\text{Conformal Osculating Barthel–Kropina}} - \text{BIC}_{\Lambda \text{CDM}}.}
$$

The interpretation follows the Jeffreys’ scale \cite{Jeffreys}:

\begin{itemize}
\item $|\Delta \text{AIC}| \leq 2$: Models are statistically comparable.
\item $4 \leq |\Delta \text{AIC}| < 10$: Considerably less support for the model.
\item $|\Delta \text{AIC}| \geq 10$: Strongly disfavored.
\item $|\Delta \text{BIC}| \leq 2$: Weak evidence against the model.
\item $2 < |\Delta \text{BIC}| \leq 6$: Moderate evidence against the model.
\item $|\Delta \text{BIC}| > 6$: Strong evidence against the model.
\end{itemize}

We also evaluate the statistical significance of the fit using the p-value
$$
p = 1 - \mathcal{F}_{\chi^2_{\text{min}}}(\chi \mid \nu),
$$
where $\mathcal{F}_{\chi^2_{\text{min}}}(\chi \mid \nu)$ is the cumulative Chi-squared distribution with $\nu$ degrees of freedom (data points minus free parameters). A p-value $p < 0.05$ indicates that the model provides a statistically significant fit. In our analysis, the degrees of freedom (DOF) is 1615 for the $\Lambda$CDM model and {1613 for the COBK model}

\section{Summary and Discussion of the Results}\label{sect3}

In the present section, we provide the detailed discussion of the results obtained in the previous sections for the {COBK model}.

\subsection{MCMC Results}
\begin{figure}[H]
  \begin{subfigure}{0.90\textwidth}
    \includegraphics[width=\linewidth]{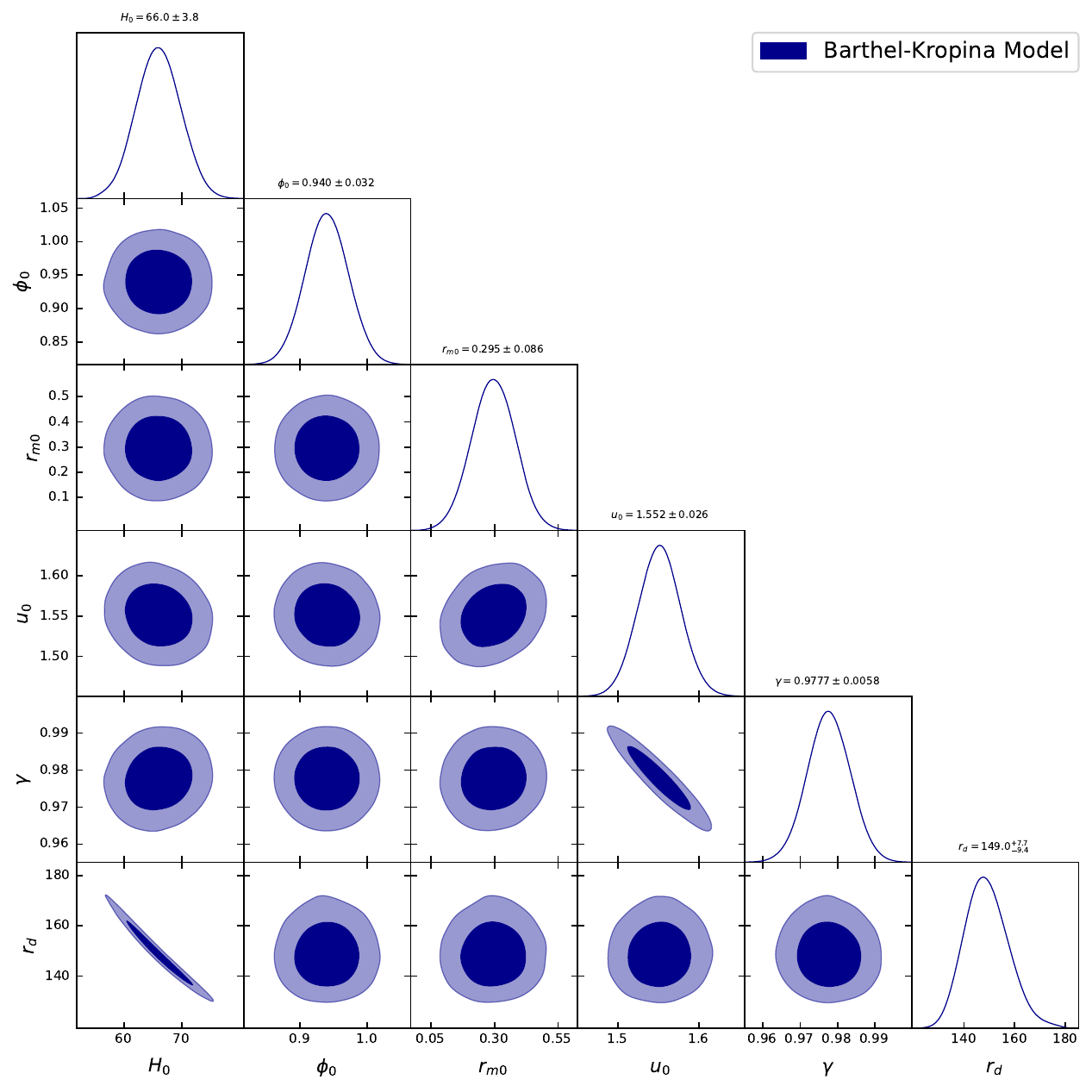}
  \end{subfigure}
\caption{The constraints 
 on the parameters of the {conformal osculating Barthel–Kropina model} using DESI DR2, SNe Ia, and CC measurements at the 68\% (1 $\sigma$) and 95\% (2 $\sigma$) confidence levels.}\label{fig_1}
\end{figure}

{Figure~\ref{fig_1} shows the corner plot for the COBK model. The diagonal panels display the 1D marginalized posterior distributions for each parameter of the COBK model, while the off-diagonal panels present the 2D marginalized confidence contours at the 68\% and 95\% confidence levels. Table~\ref{tab_1} summarizes the numerical constraints obtained for the COBK model and the $\Lambda$CDM model from the MCMC analysis.}

\unskip

\begin{table}[H]
\caption{The 
 numerical values of the parameters for the {conformal osculating Barthel–Kropina model} and of the $\Lambda$CDM model, showing mean values with 68\% credible intervals (1$\sigma$) along with the corresponding priors.}\label{tab_1}
\renewcommand{\arraystretch}{1.2} 
\renewcommand{\belowrulesep}{.1pt}
\renewcommand{\aboverulesep}{.1pt}
\begin{tabularx}{\textwidth}{cCCC}
    \toprule
    \textbf{Cosmological Models} & \textbf{Parameter} & \textbf{Prior} & \textbf{JOINT} 
 \\
    \hline
    \multirow{3}{*}{$\Lambda$CDM Model} 
     & $H_0$ & $\mathcal{U}[50,100]$ & $67.8 \pm 3.7$ \\
     & $\Omega_{m0}$ & $\mathcal{U}[0,1]$ & $0.3092 \pm 0.0086$ \\
     & $r_d$ & $\mathcal{U}[100,300]$ & $148.5 \pm 7.5$ \\
     \hline
     \multirow{6}{*}{COBK Model}
     & $H_0$ & $\mathcal{U}[50,100]$ & $66.0{\pm 3.8}$ \\
     & $\phi_{0}$ & $\mathcal{U}[0,1]$ & $0.940 \pm 0.032$ \\
     & $r_{m0}$ & $\mathcal{U}[0,1]$ & $0.295 \pm 0.086$ \\
     & $u_{0}$ & $\mathcal{U}[1,3]$ & $1.552\pm 0.026$ \\
     & $\gamma$ & $\mathcal{U}[0,2]$ & $0.977\pm 0.0058$ \\
     & $r_d$ & $\mathcal{U}[100,300]$ & $149.0_{-9.4}^{+7.7}$ \\
    \bottomrule
\end{tabularx}
\end{table}

{We first compare the predicted values of $H_0$ obtained from the $\Lambda$CDM and COBK models using the combination of the DESI DR2, PP, and CC measurements. For the $\Lambda$CDM model, we obtain $H_0 = 67.8 \pm 3.7,$ while, on the other hand, the COBK model predicts $H_0 = 66.0 \pm 3.9$. This shows a deviation of about $0.33\sigma$ from the $\Lambda$CDM prediction. The predicted value of the matter density by the COBK model shows close agreement with the value presented by the DESI DR2 $r_{m0} = 0.297 \pm 0.0086$ \cite{Karim2025}.}

{In addition, the $\Lambda$CDM model produces $r_d = 148.7 \pm 7.5$, while the COBK model predicts $r_d = 148.7^{+7.6}_{-9.4}$, showing a deviation of about $0.05\sigma$ when compared to the $\Lambda$CDM model. Also, the mean values in both cases the $\Lambda$CDM and COBK models are close to those predicted by the Planck 2018 results \cite{Aghanim2020}, ($r_d = 147.21 \pm 0.23$). Indeed, we obtain larger uncertainties since we consider only late-time datasets. Moreover, in the CC measurements, we have taken into account the full systematic and statistical uncertainties \cite{Moresco2018,Moresco2020}.}

{It is worth noting that the predicted value of $H_0$ by the COBK model shows a clear deviation from the local measurement obtained by \cite{T2} ($H_0 = 73.04 \pm 1.04$). In addition, to alleviate the Hubble tension, the COBK model would need to reduce the sound horizon by approximately $7\%$, thereby increasing the inferred value of $H_0$; however, such a reduction is not achieved in this case. Consequently, while the COBK model remains consistent with the Planck 2018 results, it does not provide a definitive solution to the Hubble tension. In the corner plot, we also see that $u_0$ and $\gamma$ have a negative correlation.}

\subsection{Hubble Parameter, and Hubble Residual Results}

\begin{figure}[H]
\begin{subfigure}{.48\textwidth}
\includegraphics[width=\linewidth]{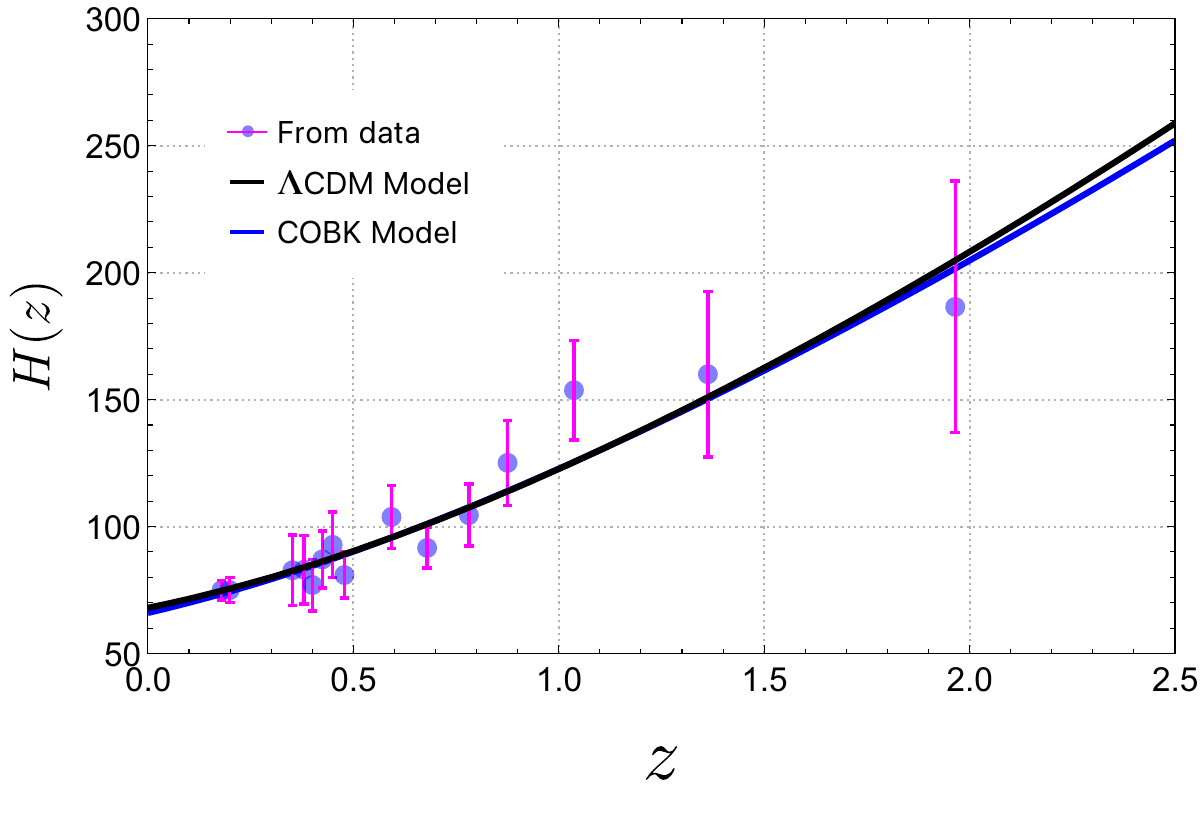}
\end{subfigure}
\hfil
\begin{subfigure}{.47\textwidth}
\includegraphics[width=\linewidth]{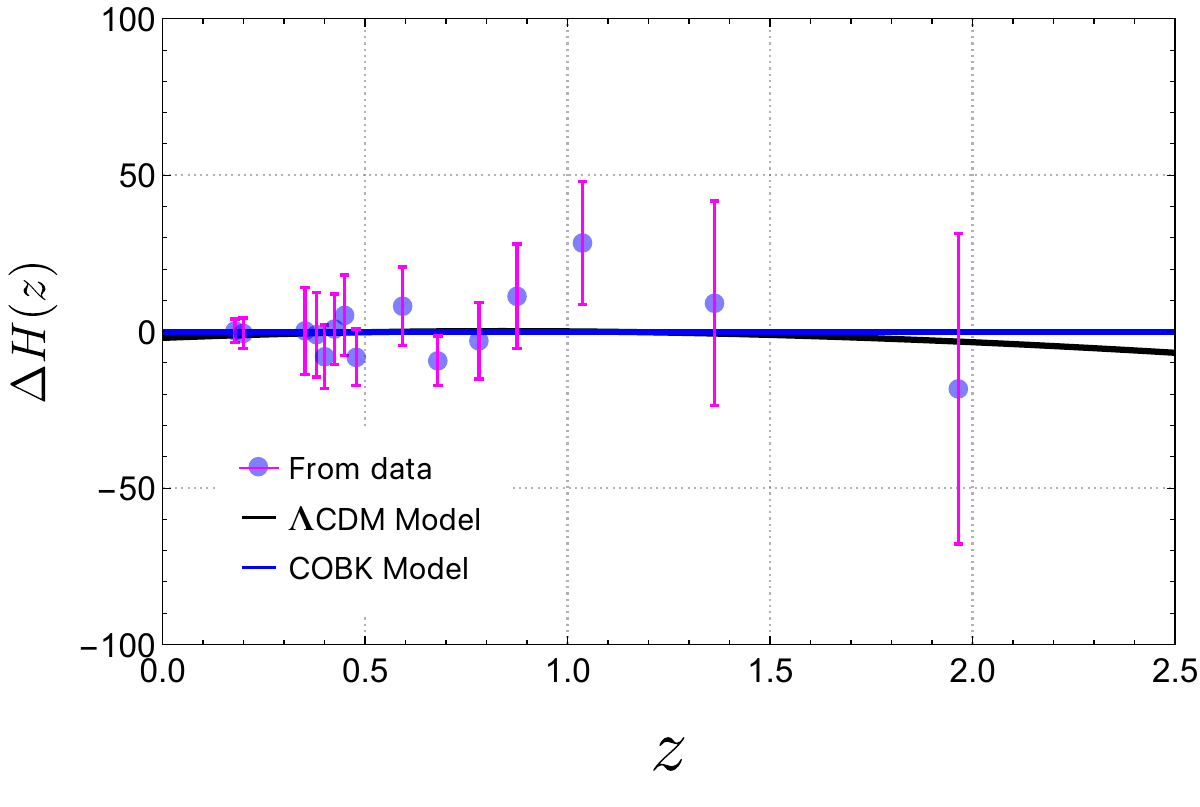}
\end{subfigure}
\caption{{The 
 comparative analysis of the conformal osculating Barthel–Kropina model against the $\Lambda$CDM model and the CC measurements, which are represented by blue dots with corresponding green error bars. The left panel shows the evolution of the Hubble function $H(z)$, while the right panel shows the evolution of the Hubble residual $\Delta H(z)$.}}\label{fig_2}
\end{figure}

{Figure~\ref{fig_2} shows the evolution of the Hubble function and Hubble residuals. The evolution of the Hubble function shows that the COBK model is in close agreement with the $\Lambda$CDM and CC measurements at $z < 1.70$, while the COBK model shows a deviation from the $\Lambda$CDM model at $z > 1.70$. A similar behaviour can be observed in the residual plots. The COBK model shows close agreement with the $\Lambda$CDM model at $z < 1.70$ and a deviation beyond $z > 1.70$.}
\vspace{-6pt}

\subsection{Cosmographic Results}

\begin{figure}[H]
\begin{subfigure}{.48\textwidth}
\includegraphics[width=\linewidth]{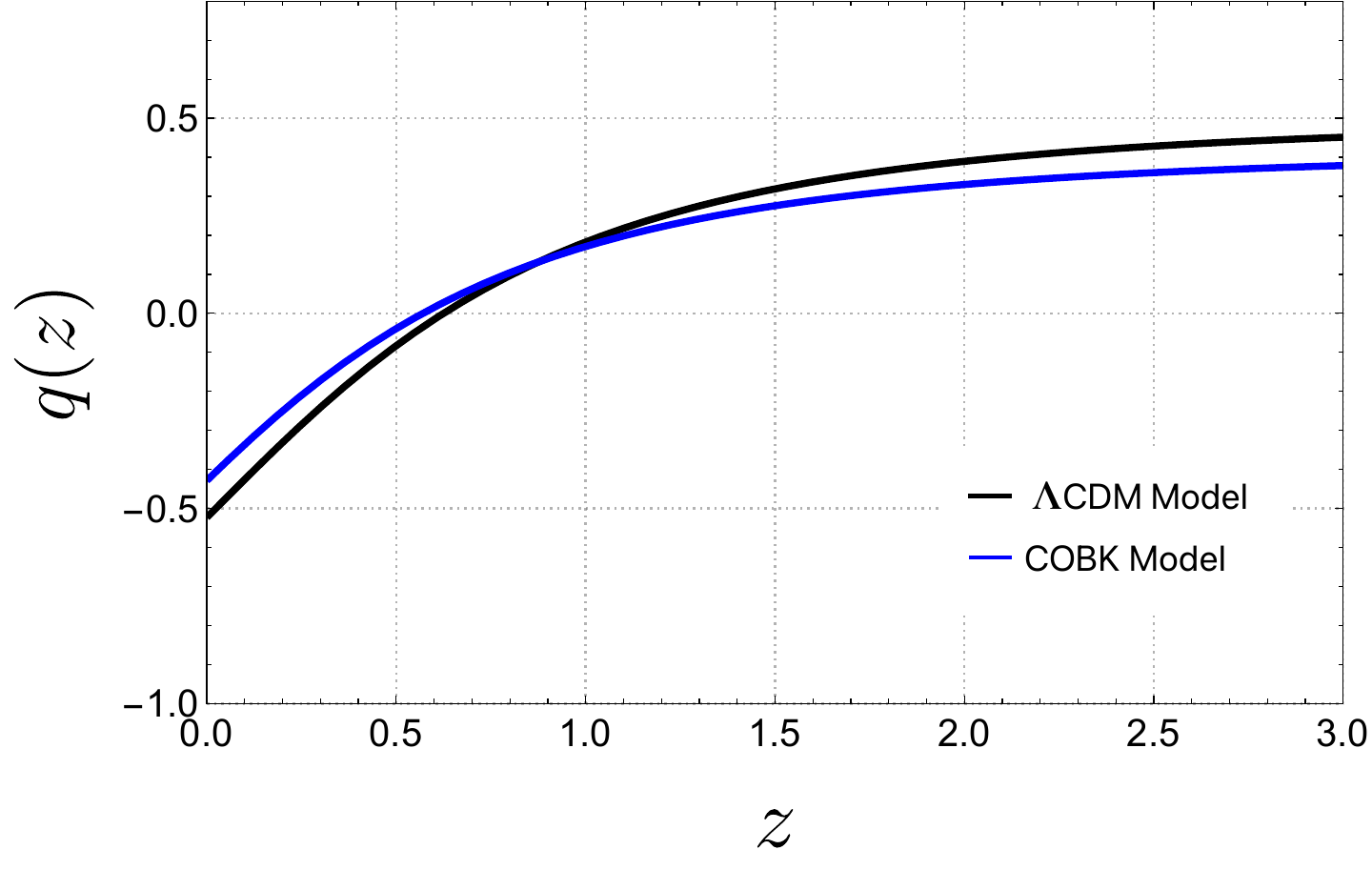}
\end{subfigure}
\hfil
\begin{subfigure}{.48\textwidth}
\includegraphics[width=\linewidth]{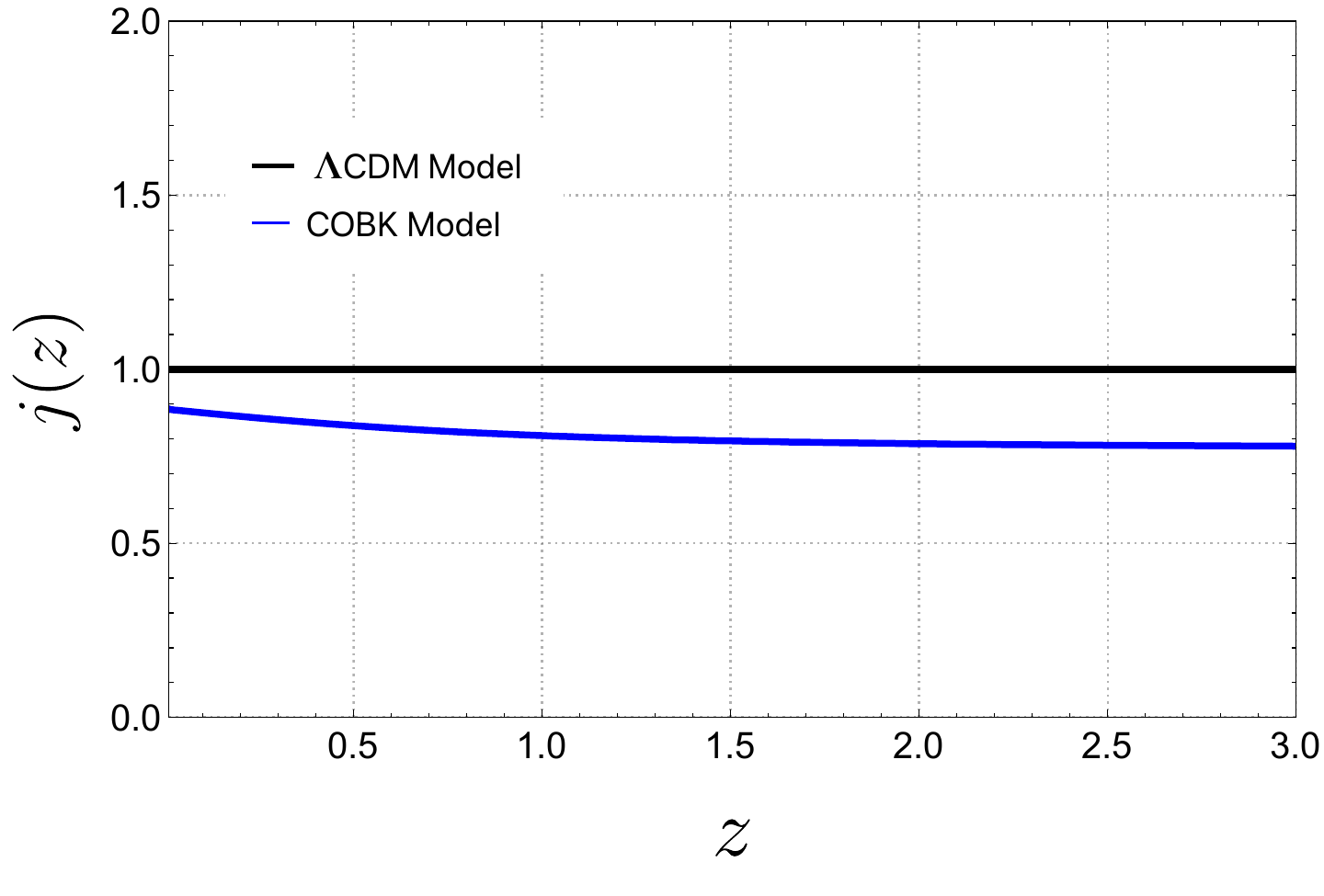}
\end{subfigure}
\caption{{The 
 evolution of the cosmographic parameters of the conformal osculating Barthel–Kropina model compared to the $\Lambda$CDM model. The deceleration parameter $q(z)$ is shown in the left panel, while the jerk parameter $j(z)$ is represented in the right panel.}}\label{fig_3}
\end{figure}

{Figure~\ref{fig_3} shows the evolution of the cosmographic parameters. The left panel presents the evolution of the deceleration parameter $q(z)$. The $\Lambda$CDM model shows a smooth transition from the decelerating to the accelerating phase, while the COBK model also shows a smooth transition from the decelerating to the accelerating phase. The $\Lambda$CDM model predicts $q_0 = -0.514$ at present, while the COBK model predicts $q_0 = -0.415$. Also, we consider the transition redshift $z_{tr}$, which indicates the redshift at which the cosmological model shows the transition from the decelerating to the accelerating phase. In the case of the $\Lambda$CDM model, the model presents $z_{tr} = 0.628$, while the COBK model predicts $z_{tr} = 0.564$.}

{The right panel shows the evolution of the jerk parameter $j(z)$. The $\Lambda$CDM model predicts $j(z) = 1$ throughout the evolution, while the COBK model shows a deviation from the $\Lambda$CDM case. Specifically, ($j(z) < 1$) for the COBK model, suggesting a slight departure from the standard $\Lambda$CDM behaviour.}

\subsection{Dimensionless Matter Density $r_m$ and Conformal Factor $\phi$}

\begin{figure}[H]
\begin{subfigure}{.48\textwidth}
\includegraphics[width=\linewidth]{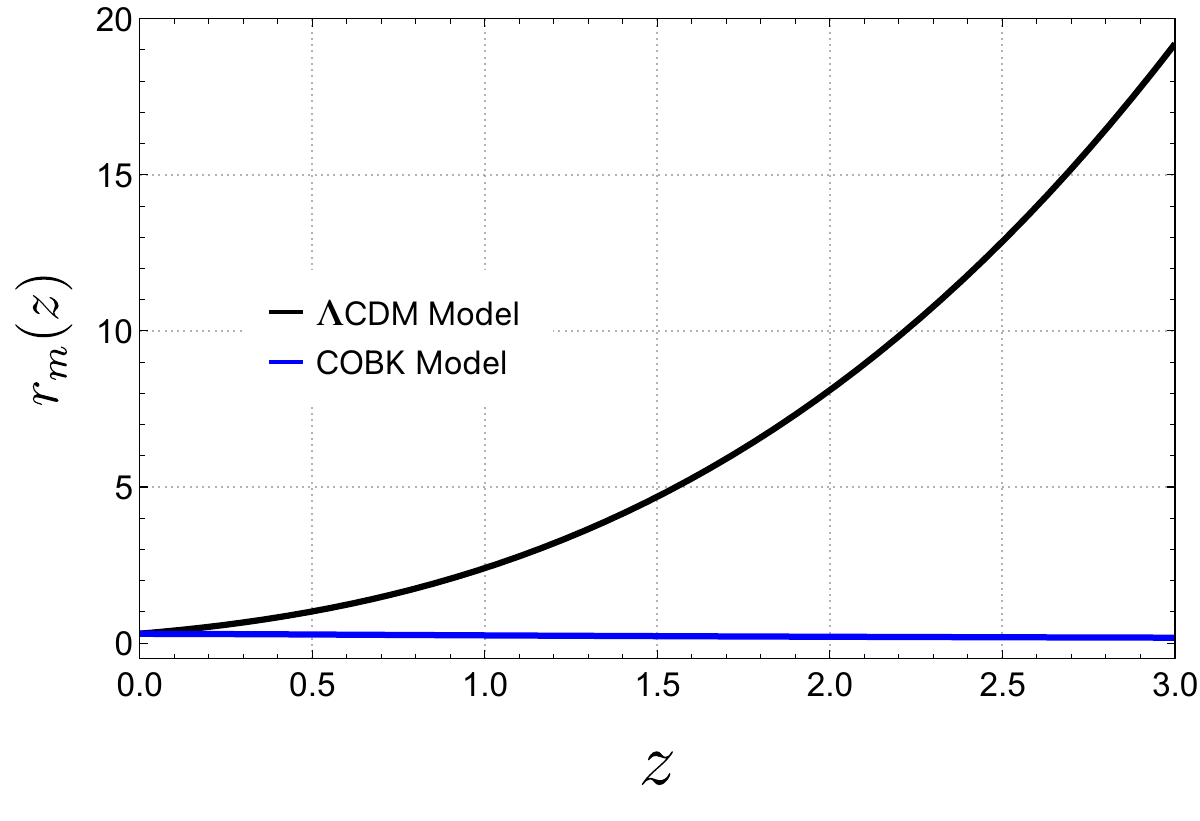}
\end{subfigure}
\hfil
\begin{subfigure}{.48\textwidth}
\includegraphics[width=\linewidth]{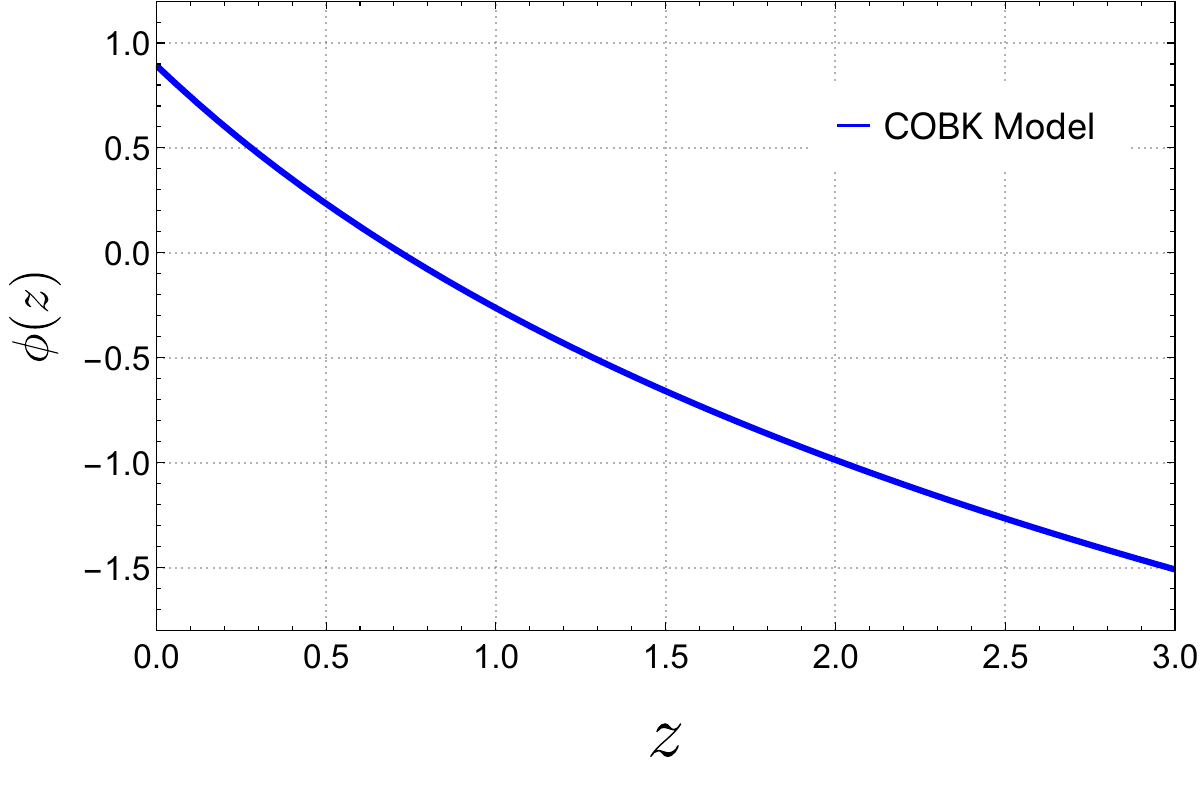}
\end{subfigure}
\caption{{The 
 evolution of the matter density in the conformal osculating Barthel–Kropina model as compared to the $\Lambda$CDM model (\textbf{left panel}),  and the evolution of the conformal factor $\phi$ (\textbf{right panel}).}}\label{fig_4}
\end{figure}

{Figure~\ref{fig_4} shows the variation of the matter density content in the $\Lambda$CDM and COBK models. In the left panel, it can be observed that both the COBK and $\Lambda$CDM models predict the same value of 0.312 at $z = 0$. However, for $z > 0.25$, the $\Lambda$CDM model predicts a higher matter density than the COBK model. The right panel shows the evolution of the conformal factor $\phi$. The conformal factor has negative values for $z > 0.69$, and positive values for $z < 0.69$.}
\vspace{-3pt}

\subsection{Evolution of Effective Energy Density and Effective Pressure of the Scalar Field}

\begin{figure}[H]
\begin{subfigure}{.48\textwidth}
\includegraphics[width=\linewidth]{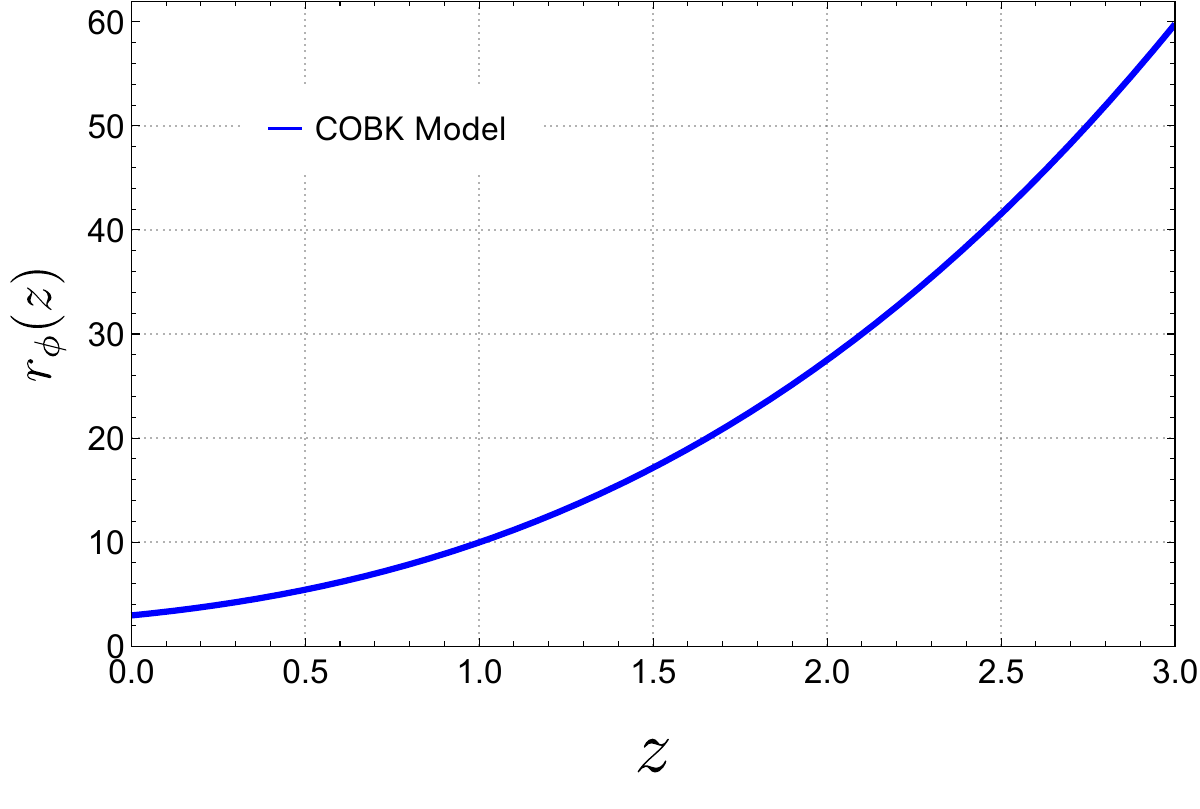}
\end{subfigure}
\hfil
\begin{subfigure}{.48\textwidth}
\includegraphics[width=\linewidth]{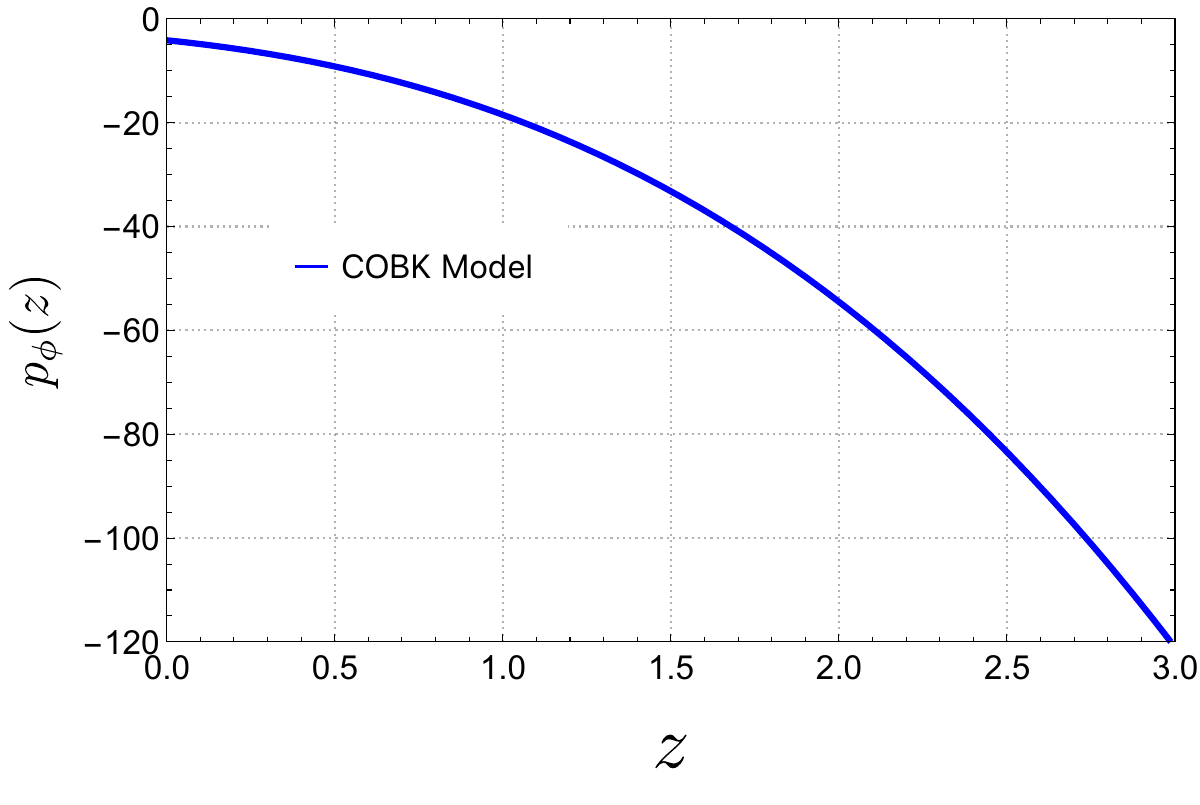}
\end{subfigure}
\caption{{The 
 evolution of the energy density (\textbf{left panel}) and the effective pressure (\textbf{right panel}) for the conformal osculating Barthel–Kropina model.}}\label{fig_5}
\end{figure}

{Figure~\ref{fig_5} shows the evolution of the effective energy density $r_\phi$ and effective pressure $p_\phi$ with respect to redshift for the COBK model. The energy density $r_{\phi}(z)$ increases as redshift increases, showing that the universe was denser at earlier times. In contrast, the effective pressure $p_{\phi}(z)$ remains negative and decreases further with redshift, indicating an increasingly strong repulsive or accelerating effect in the past.}
\vspace{-3pt}

\subsection{Statistical Results}

{Table~\ref{tab_2} presents the comparative statistical analysis between the standard $\Lambda$CDM model and the COBK model. The COBK model yields a lower total Chi-squared value ($\chi^2_{\text{tot}} = 1549.01$), and a reduced Chi-squared value of $\chi^2_{\text{red}} = 0.960$, compared to $\chi^2_{\text{red}} = 0.975$ for $\Lambda$CDM, indicating a slightly better overall fit to the observational data.}

\begin{table}[H]
\caption{The 
 statistical metrics for the $\Lambda$CDM and {conformal osculating Barthel–Kropina model}, including $\chi^2_{\text{tot}}$, $\chi^2_{\text{red}}$, AIC, $\Delta$AIC, BIC, $\Delta$BIC, p-value, and {$|\Delta \ln \mathcal{Z}_{\Lambda\mathrm{CDM}, \mathrm{COBK}}|$.}}\label{tab_2}
\renewcommand{\arraystretch}{1.1} 
\renewcommand{\belowrulesep}{.1pt}
\renewcommand{\aboverulesep}{.1pt}
\begin{adjustwidth}{-\extralength}{0cm}
\begin{tabularx}{\fulllength}{lCCCCCCCc}
   \toprule
    \textbf{Models 
} & \boldmath{$\chi^2_{\text{\textbf{tot}}}$} & \boldmath{$\chi^2_{\text{\textbf{red}}}$} & \textbf{AIC} & \boldmath{$\Delta$}\textbf{AIC} & \textbf{BIC} & \boldmath{$\Delta$}\textbf{BIC} & \textbf{\emph{p}-Value} & \boldmath{$|\Delta \ln \mathcal{Z}_{\Lambda\mathrm{\textbf{CDM}}, \mathrm{\textbf{COBK}}}|$} \\ 
    \hline
    $\Lambda$CDM Model & 1574.88 & 0.975 & 1597.04 & 0 & 1580.88 & 0 & 0.758 & 0 \\
    \hline
    COBK Model & 1549.01 & 0.960 & 1561.01 & $-$19.87 & 1593.34 & $-$3.70 & 0.867 & 14.88 \\
    \bottomrule
\end{tabularx}
\end{adjustwidth}
\end{table}

{In terms of information criteria, the COBK model achieves an AIC value of 1561.01, corresponding to $\Delta$AIC = $-19.87$, which strongly favors the COBK model over $\Lambda$CDM according to Jeffreys’ scale. However, the BIC value (1593.34) produces $\Delta$BIC = $-3.70$, suggesting only moderate evidence in favor of the COBK model, once the additional model parameters are penalized.}

{Moreover, the COBK model shows a higher p-value ($p = 0.867$) compared to $\Lambda$CDM ($p = 0.758$), indicating stronger statistical consistency with the data. The logarithmic Bayes factor difference, $|\Delta \ln \mathcal{Z}_{\Lambda\mathrm{CDM}, \mathrm{COBK}}| = 14.88$, provides decisive evidence supporting the COBK model.}

\section{Conclusions and Final Remarks}\label{sect4}

One of the fundamental assumptions, and results, of present-day physics is that the gravitational interaction can be successfully described only in geometric terms. Despite the tremendous initial success of the Riemannian geometry in this field, the recent observational results suggest that general relativity, as well as its mathematical foundation, may require a straightforward extension, and generalization. Within the multiple possibilities offered by the many existing geometric formalisms, Finsler geometry seems to be an important candidate for the reconstruction of the fundamentals of the gravitational theories. In the present work, we have investigated the cosmological implications, and tests, of a specific Finslerian type gravitational and cosmological model, the conformal Barthel--Kropina model, which is based on the extension through the introduction  of a conformal transformation of the Barthel--Kropina models \cite{B3}. 

The Barthel--Kropina-type cosmological models are based on three mathematical assumptions. The first one is the assumption the Finsler metric function, describing the properties of the gravitational interaction,  is a Kropina type $(\alpha, \beta )$ metric, with $F=\alpha ^2/\beta$.  Secondly, the osculating approach is adopted for the Finsler--Kropinsa metric, as a result of which one obtains the important result that the Finsler $g(x,y)$ is transformed into a Riemannian metric $g(x,Y(x))$. Finally, the first two assumptions imply that the connection of the Finsler--Riemann metric is the corresponding Levi--Civita connection, which, in Finsler geometry, is called the Barthel connection. With the use of this connection, one can calculate the basic geometric quantities--curvature tensors and their contractions, and write down the generalized Einstein tensor. Once the field equations are obtained, one can proceed to the investigation of their cosmological and astrophysical properties.

An extension of the Barthel--Kropina type theories can be obtained by introducing a conformal transformation of the metric function of the Finsler space. We can define a conformal transformation in the Finsler geometry in the following way. Let $F^n =\left(M^n,L\right)$ and $\tilde{F}^n = \left(Mn, \tilde{L}\right)$ be two Finsler spaces defined on the same base manifold $M^n$, where  $L$ is the Finsler metric function. If the angle between any two tangent vectors in $F^n$ is equal to the angle in $\tilde{F}^n$, then $F^n$ is called conformal to $\tilde{F}^n$. Moreover, the transformation $L\rightarrow \tilde{L}$ is called a conformal transformation of the metric \cite{B3}. A conformal transformation  implies the existence of a scalar function $\sigma  (x)$ with the property $ \tilde{L} = e^{\sigma (x)}L$. For the case of an $(\alpha, \beta)$ metric, the condition $\tilde{L} = e^{\sigma (x)}L$ is equivalent to $ \tilde {L} = e^{\sigma (x)}\alpha, e^{\sigma (x)}\beta$, which implies $\tilde{g}_{IJ} = e^{2\sigma (x)}g_{IJ}$, and $\tilde{A}_I = e^{\sigma (x)}A_I$.  

From a physical point of view, the consideration of the conformal transformations leads to the introduction in the mathematical formalism of the Barthel--Kropina geometries of a new degree of freedom, a scalar field related to the conformal transformation. This allows us to obtain an interesting cosmological model, in which by fixing the conformal factor in a simple way, one arrives to a model depending on a single independent parameter $\gamma$, which determines the components $A_I$ of the one-form $\beta $ as $A_I=\left(e^{\gamma \phi},0,0,0\right)$. In our investigations, we have considered $\gamma$ as a free parameter of the theory, and its value has been found by comparing the theoretical model with the observations.  

Once the theoretical model of the conformal Barthel--Kropina geometry is formulated, our main goal was, on one hand, to confront the model with the observations, and, on the other hand, to obtain the optimal value of the free parameter of the {COBK model. The COBK model was compared against several observational datasets, including Cosmic Chronometers, Type Ia Supernovae, and Baryon Acoustic Oscillations.} For the statistical analysis Markov Chain Monte Carlo (MCMC) methods were used. The results were also compared with the predictions of the standard $\Lambda$CDM model. 

From the comparison with the observational data, it turns out that the {COBK model is favored over the $\Lambda$CDM model by the cosmological data}. On the other hand, the conformal field $\phi(z)$ has an interesting variation, indicating an evolution from negative values at large redshifts of the order of $z=3$ to negative values at the present time. We would also like to point out that even small deviations from the value of $\gamma \approx 1$ leads to significant differences with respect to the observations. 

To conclude, our present results indicate that the conformal Barthel--Kropina cosmological models provide a satisfactory fit to the observational data, suggesting they represent a viable alternative to the standard cosmological models based on Riemannian geometry, and standard general relativity.
\vspace{6pt}



\authorcontributions{Formal analysis, R.H., T.H. and S.V.S.; investigation, R.H., T.H., S.V.S. and S.K.J.P; statistical analysis, H.C. and S.K.J.P.; writing original draft, T.H. and H.C. All the authors have substantially contributed to the present work. All authors have read and agreed to the published version of the manuscript.}
\funding{This research received no external funding.} 

\dataavailability{The original contributions presented in this study are included in the article. Further inquiries can be directed to the corresponding author.}

\conflictsofinterest{The authors declare no conflicts of interest. The funders had no role in the design of the study; in the collection, analyses, or interpretation of data; in the writing of the manuscript; or in the decision to publish the results.}

\begin{adjustwidth}{-\extralength}{0cm}

\printendnotes[custom]

\reftitle{References}

\PublishersNote{}
\end{adjustwidth}

\end{document}